\documentclass[print, amsmath,amssymb, aps,prc,nofootinbib,notitlepage,twocolumn,superscriptaddress]{revtex4-1}

\usepackage[usenames,dvipsnames]{xcolor}
\usepackage{graphicx}% Include figure files
\usepackage{dcolumn}% Align table columns on decimal point
\usepackage{bm}% bold math
\usepackage[utf8]{inputenc}
\usepackage{microtype}
\usepackage{dcolumn}
\usepackage{graphicx}
\usepackage{enumerate}

\newcommand{\nn}{\nonumber}
\renewcommand{\vec}[1]{{\bf #1}}

\newcommand{\etal}{\emph{et al.}}

\newcommand{\vnabla}{\boldsymbol{\mathbf\nabla}}

\newcommand{\nuc}[2]{$^{#1}$\textrm{#2}}

\begin{document}

\title{The impact of the surface energy coefficient
       on the deformation properties of atomic nuclei as predicted by
       Skyrme energy density functionals}

\author{W. Ryssens}
%\email{w.ryssens@ipnl.in2p3.fr}
\affiliation{IPNL, Universit\'e de Lyon, Universit\'e Lyon 1, CNRS/IN2P3, F-69622 Villeurbanne, France}

\author{M. Bender}
%\email{bender@ipnl.in2p3.fr}
\affiliation{IPNL, Universit\'e de Lyon, Universit\'e Lyon 1, CNRS/IN2P3, F-69622 Villeurbanne, France}

\author{K. Bennaceur}
%\email{bennaceur@ipnl.in2p3.fr}
\affiliation{IPNL, Universit\'e de Lyon, Universit\'e Lyon 1, CNRS/IN2P3, F-69622 Villeurbanne, France}
\affiliation{Department of Physics, PO Box 35 (YFL), FI-40014 University
of Jyv{\"a}skyl{\"a}, Finland}

\author{P.-H. Heenen}
%\email{phheenen@ulb.ac.be}
\affiliation{PNTPM, CP229, Universit\'e Libre de Bruxelles, B-1050 Bruxelles, Belgium}

\author{J. Meyer}
%\email{jmeyer@ipnl.in2p3.fr}
\affiliation{IPNL, Universit\'e de Lyon, Universit\'e Lyon 1, CNRS/IN2P3, F-69622 Villeurbanne, France}

\begin{abstract}

\begin{description}

\item[Background]
In the framework of nuclear energy density functional (EDF) methods, many
nuclear phenomena are related to the deformation of intrinsic states. Their
accurate modeling relies on the correct description of the change of nuclear
binding energy with deformation. The two most important contributions to the
deformation energy have their origin in shell effects that are correlated
to the spectrum of single-particle states, and the deformability of nuclear
matter, that can be characterized by a model-dependent surface energy
coefficient $a_{\text{surf}}$.

\item[Purpose]
With the goal of improving the global performance of nuclear EDFs through the
fine-tuning of their deformation properties, the purpose of this study
is threefold. First, to analyze the impact of systematic variations of
$a_{\text{surf}}$ on properties of nuclei; second, to identify observables
that can be safely used to narrow down the range of appropriate values of
$a_{\text{surf}}$ to be targeted in future parameter fits; third, to analyze
the interdependence of $a_{\text{surf}}$ with other properties of a nuclear
EDF.

\item[Methods]
Results for a large variety of relevant observables of deformed nuclei
obtained from self-consistent mean-field calculations with a set of
purpose-built SLy5sX parametrizations of the Skyrme EDF are correlated with
the value of $a_{\text{surf}}$.

\item[Results]
The performance of the SLy5sX parametrizations for
characteristic energies of the fission barriers of $^{180}$Hg, $^{226}$Ra,
and  $^{240}$Pu, excitation energies, electromagnetic moments and moments of
inertia of superdeformed states in the $A \approx 190$ region, properties of
shape coexisting states at normal deformation in the Pb, Kr, and Zr region,
properties of octupole-deformed $^{144}$Ba, even-even Th isotopes, and
$^{110}$Zr, separation energies along isotopic and isotonic chains are
compared with available experimental data.

\item[Conclusions]
The three main conclusions are
that there is an evident preference for a comparatively low value of
$a_{\text{surf}}$, as expected from the performance of existing
parametrizations;
that the isospin dependence of the surface energy also needs further
fine-tuning in order to describe trends across the chart of nuclei; and
that a satisfying simultaneous description of fission barriers and
superdeformed states requires a better description of the single-particle
spectra.

\end{description}
\end{abstract}

\date{28 February 2019}

\maketitle

%============================================================================

\section{Introduction}

Self-consistent mean-field models based on energy density functionals
(EDF)~\cite{Bender03} are among the tools of choice to study nuclear
structure across the entirety of the nuclear chart.
Many different types of EDFs are used that are either non-relativistic or
relativistic, use contact terms with gradients or have finite-range terms
of various kinds, and use different types of density dependencies.
Because of its
flexibility and computational simplicity, however, the local non-relativistic
Skyrme EDF is arguably the most widely used form for such calculations.
Marrying a microscopic description of the nucleus with a modest computational
cost, all types of EDFs allow for the description of many
properties of nuclei, from the properties of infinite nuclear matter to
those of the ground states of finite nuclei such as their binding energy,
shell structure, radii and other characteristics of their density distribution,
to more subtle characteristics of excited states such as shape coexistence
and various types of rotational bands, to their response properties,
fission barriers and behavior in low-energy nuclear reactions.

In one way or the other, almost all of these nuclear structure phenomena are
associated with deformed intrinsic shapes of the nucleus. In the simple
picture of the nucleus as a liquid-drop, however, the ground states of all
nuclei up to charge numbers $Z$ of about 100 are spherical. Indeed, within
that framework the
deformation  of the nuclear shape leads to a decrease in total binding energy
that is mainly determined by the interplay between two contributions that
overall reduce the total binding energy: On the one hand, the surface energy
grows with deformation -- as it increases the size of the nuclear surface --
while on the other hand the Coulomb energy decreases with deformation,
as the average distance between protons, which repel each other, becomes
larger. In light systems the competition of these two terms is dominated
by the surface energy, but
with increasing $Z$ the Coulomb energy is taking over until for $Z \simeq 100$
it decreases quicker with deformation than the surface energy increases.
As a result, the spherical shape becomes a maximum of the energy landscape
which is monotonically falling off until the nucleus splits into two
or even more fragments.
In this picture, deformed states of nuclei are generated by shell effects
that give an additional contribution to the binding energy. Its quick
variation with deformation generates local minima and barriers on top of
the smooth surface from the liquid-drop energy~\cite{Brack72a,Nilsson95a}.
As has been demonstrated already a long time ago \cite{Brack75a,Brack81a},
the results of nuclear self-consistent EDF calculations can be interpreted
with the same vocabulary of macroscopic liquid-drop and microscopic
shell-correction energy, although neither of the two is
actually calculated as an ingredient of the model.

It is well-known that the dominant contribution to the surface energy
$E_{\text{surf}}$ of a spherical nucleus simply scales with $A^{2/3}$. In the
liquid-drop model (LDM), the proportionality factor between the two is called
the surface energy coefficient $a_{\text{surf}}$. Like the volume energy, the
surface energy depends on the asymmetry between proton and neutron number,
which is parameterized through a surface symmetry energy with corresponding
coefficient $a_{\text{ssym}}$. Realistic finite nuclei, however, are too small
to unambiguously separate the surface energy and its isospin dependence from
higher-order and pairing contributions to the total liquid-drop binding
energy \cite{Reinhard06a,Dobaczewski14a,Jodon16}, which compromises their
determination from experimental data for binding energies.
These coefficients, however, can also be used to characterize nuclear EDFs.
Their values can for example be calculated for the idealized model system of
semi-infinite nuclear matter. The precise value of $E_{\text{surf}}$
obtained from such calculation nonetheless still depends on choices made for
details of the modelling, see Refs.~\cite{Jodon16,JodonThesis} and
references therein.
As a consequence, it appears to be impossible to establish a unique
model-independent empirical values for $a_{\text{surf}}$ and $a_{\text{ssym}}$.

While there is an obvious correlation between the values of $a_{\text{surf}}$
and $a_{\text{ssym}}$ of a nuclear EDF and the systematics of calculated
deformation properties \cite{Bartel82,Bender99,Bender04a,Nikolov11a}, its
analysis is often compromised by the use of parametrizations that have been
adjusted with different protocols such that the parametrizations differ
in many respects, not just the surface properties. The recent series of eight
parametrizations of the standard Skyrme EDF, SLy5s1--SLy5s8
\cite{Jodon16,JodonThesis} with their systematically varied
$a_{\text{surf}}$ offer the possibility for a much cleaner separation of
the surface energy from other contributions.

The fit of the SLy5sX parametrizations is part of the ongoing efforts to
improve the descriptive power of nuclear energy density functionals, and
which concern both their functional form and the procedure used to adjust
their parameters. Important recent milestones for the latter concern
strategies to avoid unphysical finite-size instabilities \cite{Pastore13}
and the quantification of correlations between model parameters and data,
that also allows for estimating the statistical errors of observables
related to the fit protocol \cite{unedf0,unedf1,unedf2,Dobaczewski14a}.

The construction of the SLy5sX parametrizations is the first step
towards establishing a protocol to better constrain the deformation
properties of heavy nuclei such as fission barriers during
the parameter adjustment in a computationally efficient way.
To that end, the isoscalar surface energy coefficient has been varied in
small equidistant steps in the region where it can be expected to find
a realistic value. Indeed, there is no possibility to establish a unique
model-independent empirical value for $a_{\text{surf}}$ that can be
determined \textit{a priori}. Hence, one has to choose a scheme for its
calculation, carry out a series of fits that cover the relevant region
and determine the value that corresponds to a realistic description of
nuclei \textit{a posteriori}. If necessary, this value can be fed back
into a series of refits covering its vicinity until a best fit is achieved
iteratively. The fine-tuning of the D1S parametrization of the Gogny
interaction to fission properties was in fact based on similar ideas
\cite{Berger89}.

Other previous attempts to fine-tune deformation properties during an EDF's
parameter fit either relied on semiclassical estimates for the fission
barrier height, as done for SkM* in Ref.~\cite{Bartel82}, or on the adjustment
of the excitation energy of the fission isomer in the actinide region as
done for UNEDF1 and UNEDF2 in Refs.~\cite{unedf1,unedf2}.

Using the SLy5sX parametrizations as the starting point, the goal of the
present article is threefold.
\begin{enumerate}[(i)]
\item
First, we want to benchmark the descriptive power of the series of SLy5sX
parametrizations on typical properties of nuclei, scrutinizing their
differences in dependence of their value for $a_{\text{surf}}$ for
typical observables of deformed nuclei frequently calculated with
nuclear EDF methods.

\item
Second, we want to identify observables that are directly affected
by a change in $a_{\text{surf}}$ and which can then in the future be used
for the adjustment of a ``best value'' for $a_{\text{surf}}$ during a
parameter fit aiming at a unique best fit of an EDF.

\item
Third, we want to analyze to which extent a constraint on $a_{\text{surf}}$
is independent from other data, be they used during the fit or not.
As many terms in the EDF contribute to it, there is the possibility that
setting $a_{\text{surf}}$ to some specific value substantially degrades
other properties of the parametrizations that are less strictly
constrained.
\end{enumerate}
This paper is organized as follows: Section~\ref{sect:Theo} provides
a discussion of the nuclear matter properties of the SLy5sX parametrizations
of the Skyrme EDF. Section~\ref{sec:LDM} discusses the mapping of EDF
results on a liquid-drop model, which will be used as a diagnostic tool
later on. Section~\ref{sec:results} analyzes the differences of results
obtained with the set of SLy5sX parametrizations for fission barriers of
selected representative heavy and superheavy nuclei, superdeformed states
in the $A \approx 190$ mass region, shape coexisting states
at normal deformation for the example of $^{186}$Pb, $^{74}$Kr, and $^{110}$Zr,
and some selected octupole-deformed ground states. Finally,
in Sec.~\ref{sec:conclusion} we summarize the main results of the paper and
we outline new constraints for the
construction of EDFs.

%
%=========================================================================
%
\section{SL\lowercase{y5s}X parametrizations}
\label{sect:Theo}

\subsection{Energy density functional}

In the context of the Skyrme EDF method, it is customary to split
the energy density functional into five terms~\cite{Bender03}
\begin{equation}
\label{eq:Etot}
E_{\text{tot}}
  =     E_{\text{kin}}
      + E_{\text{Skyrme}}
      + E_{\text{Coul}}
      + E_{\text{pair}}
      + E_{\text{corr}}
\, ,
\end{equation}
which correspond to the kinetic energy, the actual Skyrme EDF that models of
the strong interaction between the nucleons in the particle-hole channel, the
Coulomb energy resulting from the electromagnetic repulsion between protons,
a  pairing EDF modelling the strong-interaction in the particle-particle
channel, and correction terms for spurious zero-point motion that result
from the mean-field approximation.

The SLy5sX parametrizations considered throughout this article
use the standard form of the Skyrme EDF combining central and spin-orbit
terms up to next-to-leading order in derivatives with a simple density
dependence of the gradientless terms. As for SLy5 \cite{Cha98}, the
contribution of the central interaction to the tensor terms that are
bilinear in spin-current tensor density $J_{\mu \nu}(\vec{r})$ is kept,
while the correction terms for spurious zero-point motion are limited
to the one-body contribution to the center-of-mass correction $E_{\text{cm}}$.
No additional independent tensor force is taken into account when generating
the EDF. Also, like for SLy5, the direct Coulomb
term is calculated from the point-proton density, while the Coulomb exchange
term is approximated by the local Slater approximation.

For time-reversal-invariant systems, the Skyrme EDF then takes the form
\begin{eqnarray}
\label{eq:skyrme:energy}
E_{\text{Skyrme}}^{\text{even}}
& = &   E_{\rho^2}
      + E_{\rho^{2 + \alpha}}
      + E_{\rho \tau}
      + E_{\rho \Delta \rho}
      + E_{\rho \nabla J}
      + E_{JJ}
      \nonumber \\
& = & \sum_{t=0,1} \int \! d^3r \,
      \Big[
               C^{\rho\rho}_t \, \rho_t^2 (\vec{r})
             + C^{\rho\rho\rho^{\alpha}}_t \rho_0^\alpha (\vec{r}) \, \rho_t^2 (\vec{r})
      \nonumber \\
&   &
             + C^{\rho\tau}_t \, \rho_t (\vec{r}) \, \tau_t (\vec{r})
             + C^{\rho \Delta \rho}_t \, \rho_t (\vec{r}) \, \Delta \rho_t (\vec{r})
      \nonumber \\
&   &
             + C^{\rho \nabla \cdot J}_t  \, \rho_t (\vec{r}) \, \vnabla \cdot \vec{J}_t (\vec{r})
      \nonumber \\
&   &
             - C^{sT}_t \sum_{\mu, \nu} J_{t, \mu \nu} (\vec{r}) \, J_{t, \mu \nu} (\vec{r})
      \Big] \, .
\end{eqnarray}
It is a functional of the isoscalar ($t=0$) and isovector ($t=1$)
local density $\rho_t(\vec{r})$, kinetic density $\tau_t(\vec{r})$ and
spin-current tensor density $J_{t,\mu\nu}(\vec{r})$. The latter has
nine independent Cartesian
components labeled by $\mu$ and $\nu$, with the spin-orbit current
$\vec{J}_t(\vec{r})$ being its rank-1 contraction.
The $C$ coefficients are the coupling constants of the various
terms in the isoscalar ($t=0$) and isovector ($t=1$) channels.
For further details and the definition of these quantities see
Refs.~\cite{Bender09a,Hellemans12,Ryssens15a}.

All densities entering Eq.~\eqref{eq:skyrme:energy} are even under
time-reversal. For the calculation of the rotational bands discussed
in Sect.~\ref{sect:SD:rot}, where time-reversal invariance is broken,
additional terms have to be considered that depend on the time-odd
spin density $\vec{s}_t(\vec{r})$, current density $\vec{j}_t(\vec{r})$
and kinetic spin density $\vec{T}_t(\vec{r})$ \cite{Hellemans12},
\begin{eqnarray}
\label{eq:SkTodd}
E_{\text{Skyrme}}^{\text{odd}}
& = & \sum_{t=0,1}
      \int d^3 r \,
      \Big[ C^{ss}_t \, \vec{s}_t^2 (\vec{r})
          + C^{ss\rho^\alpha}_t \, \rho_0^\alpha (\vec{r}) \, \vec{s}_t^2  (\vec{r})
      \nonumber \\
&   &
      + C^{sT}_t \, \vec{s}_t (\vec{r}) \cdot \vec{T}_t (\vec{r})
      + C^{s \Delta \vec{s}} \, \vec{s}_t (\vec{r}) \cdot \Delta \vec{s}_t (\vec{r})      \nonumber \\
&   &
      - C_t^{\rho\tau} \, \vec{j}_t^2 (\vec{r})
      + C_t^{\rho \nabla J} \, \vec{s}_t (\vec{r}) \cdot \vnabla \times \vec{j}_t (\vec{r})
      \Big]
\, .
\end{eqnarray}
This part of the EDF is colloquially called the `time-odd' part of the
functional. Although constructed out of time-odd densities, the EDF itself
is time-even. Note that some of the coupling constants in the
time-even~\eqref{eq:skyrme:energy} and time-odd~\eqref{eq:SkTodd} parts
are necessarily equal, up to a sign, for reasons of Galilean
invariance \cite{Engel75a}.
The coupling constants of the other terms in the time-odd part of the EDF
can be linked to those of the time-even part~\eqref{eq:skyrme:energy} by
calculating the entire Skyrme EDF as the expectation value of a
density-dependent zero-range two-body interaction for a Slater determinant
\cite{Engel75a,Hellemans12}.
This, however, is not always done, in particular because the term
containing the Laplacian of the spin density $\Delta \vec{s}_t (\vec{r})$
can be the source of a non-physical finite-size instability in the spin
channels~\cite{Schunck10a,Pototzky10a,Hellemans12,Pastore15} when keeping
its coupling constant at the Skyrme-force value. Adding the constraint
proposed in Ref.~\cite{Pastore13} to the fit protocol, it has been ensured
that the SLy5sX parametrizations are free of such instabilities for values
of the densities encountered in finite nuclei.

The SLy5sX parametrizations were adjusted to properties of doubly-magic
nuclei for which pairing correlations vanish at the mean-field level. The
calculations that are presented here require the introduction of pairing
correlations, which is done by solving the HFB equations with the two-basis
method~\cite{Gall1994}. We use a simple pairing EDF of the form
\begin{equation}
\label{eq:pair:edf}
\mathcal{E}_{\text{pairing}}
= \sum_{q=p,n} \frac{V_q}{4} \int d^3 r \,
  \left[ 1 - \frac{\rho_0(\vec{r})}{\rho_c}\right] \,
  \tilde{\rho}^*_q(\vec{r}) \, \tilde{\rho}_q(\vec{r}) \, ,
\end{equation}
where the $\tilde{\rho}_q(\vec{r})$ are local pairing densities that become
complex when time-reversal symmetry is broken. As all SLy5sX
parametrizations have almost the same effective mass as the SLy4
parametrization, we  have taken the same values
$V_q =-1250$~MeV~fm$^{-3}$ and $\rho_q = 0.16$~fm$^{-3}$, originally
adjusted to moments of inertia of superdeformed rotational bands in
the $A \approx 190$ region~\cite{Rigollet1999}, as done in many previous
studies using SLy4. A smooth cutoff above and below the Fermi energy is
introduced  by multiplying the contribution from the single-particle state
with index $k$ by the factor
\begin{equation}
\label{eq:pair:cutoff:2}
f_k
= \big[ 1 + \text{e}^{(\epsilon_k' - \Delta \epsilon_q)/\mu_q}
  \big]^{-1/2}
  \big[ 1 + \text{e}^{(\epsilon_k' + \Delta \epsilon_q)/\mu_q}
  \big]^{-1/2} \, ,
\end{equation}
when summing the pair densities in the basis that diagonalizes the
single-particle Hamiltonian $\hat{h}$. The cutoff depends on the distance
$\epsilon_k' \equiv \epsilon_k - \lambda_q$ of a given eigenvalue
$\epsilon_k$ of $\hat{h}$ from the Fermi energy $\lambda_q$ of the nucleon
species $q$. For the parameters, we choose $\mu_q = 0.5 \, \text{MeV}$, and
$\Delta \epsilon_q = 5.0 \, \text{MeV}$ for both protons and neutrons
as done in the past~\cite{Rigollet1999}.

Finally, unless explicitly mentioned otherwise, we have employed the
Lipkin-Nogami prescription as in Ref.~\cite{Rigollet1999} in order to
avoid a collapse of pairing correlations.

%------------------------------------------------------------------------

\subsection{Global properties of the SL\lowercase{y5s}X parametrizations}
\label{sec:differences}
%
%=============================================================================
%
\begin{table}[t!]
\caption{\label{tab:INM}
Properties of infinite nuclear matter as
obtained with the SLy5sX parametrizations:
saturation density $\rho_{\text{sat}}$ in fm$^{-3}$,
energy per particle $E/A = a_{\text{vol}}$ in MeV,
incompressibility $K_\infty$ in MeV,
isoscalar effective mass $m^*_0/m$,
symmetry energy coefficient $J = a_{\text{sym}}$ and its slope $L$ in MeV,
and enhancement factor of the Thomas-Reiche-Kuhn sum rule $\kappa_v$.
}
\begin{center}
\begin{tabular}{lcccccccc}
\noalign{\smallskip} \hline \noalign{\smallskip}
        & $\rho_{\text{sat}}$ & $E/A$ & $K_\infty$ & $m_0^*/m$ & $J$ & $L$ & $\kappa_v$ \\
\noalign{\smallskip}\hline\noalign{\smallskip}
 SLy5s1 & 0.1598 & $-15.772$ & 222.1 & 0.7392 & 31.43 & 48.1 & 0.3047 \\
 SLy5s2 & 0.1603 & $-15.818$ & 223.2 & 0.7350 & 31.60 & 48.3 & 0.3063 \\
 SLy5s3 & 0.1607 & $-15.864$ & 224.3 & 0.7309 & 31.77 & 48.4 & 0.3082 \\
 SLy5s4 & 0.1612 & $-15.911$ & 225.4 & 0.7273 & 31.94 & 48.5 & 0.3105 \\
 SLy5s5 & 0.1618 & $-15.958$ & 226.4 & 0.7243 & 32.11 & 48.6 & 0.3131 \\
 SLy5s6 & 0.1623 & $-16.005$ & 227.3 & 0.7217 & 32.29 & 48.8 & 0.3160 \\
 SLy5s7 & 0.1629 & $-16.053$ & 228.3 & 0.7196 & 32.46 & 48.9 & 0.3191 \\
 SLy5s8 & 0.1634 & $-16.101$ & 229.1 & 0.7178 & 32.64 & 49.0 & 0.3225 \\
\noalign{\smallskip} \hline
\end{tabular}
\end{center}
\end{table}
%
%===================
%

The SLy5sX parametrizations were adjusted with an auxiliary condition on
their surface energy coefficient in such a way that it takes a different
value for each of them.
The fit protocol used for their adjustment~\cite{Jodon16} is an update of
the Saclay-Lyon protocol originally set up in the 1990s~\cite{Cha98}.
The most noteworthy  differences to Ref.~\cite{Cha98} are the additional
constraint proposed in Ref.~\cite{Pastore13} that prevents the often
encountered appearance of unphysical finite-size instabilities in the
spin-channels already mentioned above, and an additional constraint on
the slope of the symmetry energy that will be commented on below.

Tables~\ref{tab:INM} and~\ref{tab:parameters} list the most relevant
properties of the SLy5sX parametrizations for the model systems of
infinite (INM) and semi-infinite (SINM) nuclear matter, respectively.
While the properties of INM listed in Tab.~\ref{tab:INM} can be obtained
from simple functions of the parameters, the properties of SINM have to be
deduced from a numerical calculation of this system. The latter
can be carried out in different frameworks, each of which yields slightly
different values, which is the reason why Tab.~\ref{tab:parameters} lists
values for $a_{\rm surf}$ calculated from quantal
Hartree-Fock (HF) as well as semi-classical Extended Thomas-Fermi (ETF)
and Modified Thomas-Fermi (MTF) calculations~\cite{Jodon16,JodonThesis}.
While values for $a_{\rm surf}$ obtained with different models for SINM are
visibly different for a given parametrization, the difference between values
of different parametrizations obtained within the same model for SINM
is almost independent on the choice of model~\cite{Jodon16}.
We also list values for the surface symmetry energy coefficient
$a_{\rm ssym}$, but only from HF calculations. Establishing a value for
$a_{\rm ssym}$ turns out to be even more
delicate than determining $a_{\rm surf}$. The generalization of MTF to
asymmetric SINM is not straightforward and requires further approximations,
and there are several possibilities for the protocol to extract $a_{\rm ssym}$
from a series of SINM calculations with varying asymmetry. A detailed
analysis of the model dependence of the resulting $a_{\rm ssym}$ from such
calculations will be presented elsewhere~\cite{Meyer18x}.

%
%===================
%
\begin{table}[t!]
\caption{\label{tab:parameters}
Properties of semi-infinite nuclear matter as obtained with the SLy5sX
parametrizations:
Surface energy coefficient $a_{\rm surf}$ as obtained within the HF, ETF
and MTF approaches and
surface symmetry energy coefficient $a_{\rm ssym}$ as obtained within the
HF method, all in MeV. The last column lists the isoscalar coupling
constant $C^{\nabla \cdot J}_{0} = 3 C^{\nabla \cdot J}_{1}$  of the
spin-orbit term in the EDF~\eqref{eq:skyrme:energy} in MeV~fm$^{-5}$.
}
\begin{tabular}{lccccc}
\hline
\noalign{\smallskip}
 & $a^{\rm (MTF)}_{\rm surf}$
 & $a^{\rm (ETF)}_{\rm surf}$
 & $a^{\rm (HF)}_{\rm surf}$
 & $a_{\rm ssym}^{\rm (HF)}$
 & $C^{\rho \nabla \cdot J}_{0}$  \\
\noalign{\smallskip}
\hline
\hline
\noalign{\smallskip}
SLy5s1 & 18.00 & 17.15 & 17.55 & $-48.09$ & $-86.61$  \\
SLy5s2 & 18.20 & 17.34 & 17.74 & $-48.21$ & $-85.71$  \\
SLy5s3 & 18.40 & 17.53 & 17.93 & $-48.56$ & $-84.65$  \\
SLy5s4 & 18.60 & 17.73 & 18.12 & $-49.01$ & $-83.50$  \\
SLy5s5 & 18.80 & 17.92 & 18.31 & $-49.73$ & $-82.31$  \\
SLy5s6 & 19.00 & 18.11 & 18.50 & $-50.53$ & $-81.08$  \\
SLy5s7 & 19.20 & 18.31 & 18.70 & $-51.58$ & $-79.82$  \\
SLy5s8 & 19.40 & 18.50 & 18.89 & $-52.70$ & $-78.54$  \\
\noalign{\smallskip}
\hline
\end{tabular}
\end{table}
%
%===================
%

The SLy5sX parametrizations were adjusted with a constraint on the
MTF value of $a_{\text{surf}}$, which is the most computationally-friendly
approach for its calculation. Going from SLy5s1 with the lowest
$a_{\text{surf}}^{\text{MTF}} = 18.0 \; \text{MeV}$ to SLy5s8 with the
highest value of $19.4 \; \text{MeV}$ in equal steps of
$0.2 \; \text{MeV}$ covers the range typically found for widely-used
Skyrme parametrizations~\cite{Jodon16}.

Because of the limited number of degrees of freedom of the Skyrme EDF,
the value of $a_{\text{surf}}$ cannot be varied independently from
the other nuclear matter properties. As with $a_{\text{surf}}$ one
property is constrained to a precise value, the others readjust
themselves. As can be seen from Tab.~\ref{tab:INM}, the properties
of INM vary slowly and systematically as a function of $a_{\rm surf}$.
In fact, all nuclear matter properties listed in Table~\ref{tab:INM} have
been constrained in one way or the other during the parameter fit.
Keeping the values for $\rho_{\text{sat}}$,
$E/A = a_{\text{vol}}$, $K_\infty$, $J = a_{\text{sym}}$, $m^*_0/m$,
and $\kappa_v$ near the empirical ones has already been proposed in the
original fit protocol of \cite{Cha98}. The slope of the symmetry
energy $L$, which is not well fixed by data on finite nuclei, was not
constrained in the protocol of Ref.~\cite{Cha98}. During the parameter fit
of the SLy5sX, however, the value of $L$ started to change on a large scale
when varying $a_{\text{surf}}$, such that it has been constrained to the
interval of $(50 \pm 2) \, \text{MeV}$ in order to keep bulk properties
at similar values~\cite{Jodon16}.
The variation of these nuclear matter properties when going from one
SLy5sX parametrization to another can be expected to have some impact
on the properties of finite nuclei.

%
%===================
%
\begin{figure}[t!]
\includegraphics[width=8.2cm]{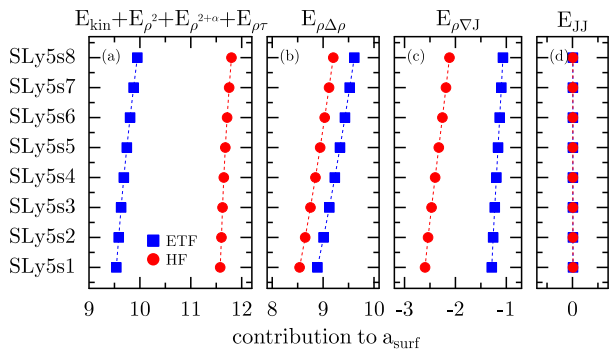}
\caption{\label{fig:esurf:decomp}
Decomposition of the surface energy coefficient in the contributions from
terms in the EDF that contribute to INM [panel (a)], gradient terms
[panel (b)], spin-orbit terms [panel (c)] and tensor terms [panel (d)].
All panels share the same energy scale.
}
\end{figure}
%
%==================
%

The value for $a_{\rm surf}$ is determined by all isoscalar terms in the
time-even part of the Skyrme EDF, Eq.~\eqref{eq:skyrme:energy}, as well as the kinetic
energy $E_{\text{kin}}$ in Eq.~\eqref{eq:Etot}.
Figure~\ref{fig:esurf:decomp} shows the decomposition of $a_{\text{surf}}$
into terms that contribute to $E/A$ of INM
($E_{\text{kin}} + E_{\rho^{2}} + E_{\rho^{2 + \alpha}} + E_{\rho \tau}$),
gradient terms in the density ($E_{\rho \Delta \rho}$), spin-orbit
terms ($E_{\rho \nabla J}$) and tensor terms ($E_{JJ}$). Results are shown
for calculations performed within the semi-classical ETF and the microscopic
HF method. For the MTF method, which because of its numerical efficiency
was the tool of choice to constrain the $a_{\text{surf}}$ of the SLy5sX
parametrizations during their fit \cite{Jodon16}, such decomposition cannot
be made.

There are obvious differences between the decompositions of
$a_{\text{surf}}$ when it is calculated with either the HF or ETF method.
We focus, however, on the decomposition of the ETF result first.
In this case, the INM and gradient
terms are of comparable size and clearly dominate, while the
spin-orbit term $E_{\rho \nabla J}$ brings a small, but non-negligible,
correction of opposite sign. For the SLy5sX parametrizations that only
have a contribution from the central force to $E_{JJ}$, but no explicit
tensor interaction, with about 10~keV this term's contribution to
$a_{\text{surf}}$ is so small that it cannot be resolved on the Figure.
All terms contribute coherently to the increase in surface energy when going
from SLy5s1 to SLy5s8, with changes in the gradient and spin-orbit terms
being larger than those of the INM terms.

When calculated in HF, however,
the contribution from INM terms is about 2~MeV larger than for ETF, whereas
the gradient and spin-orbit terms are smaller, leading to the net difference
of about 0.4~MeV between $a^{\rm (HF)}_{\rm surf}$ and
$a^{\rm (ETF)}_{\rm surf}$ reported
in Table~\ref{tab:parameters}. The slopes of the INM and spin-orbit
terms are also different when calculated in HF or ETF, but in the opposite
direction such that the slope of their sums is almost identical for
both methods as already pointed out in Ref.~\cite{Jodon16}.
The variational calculation of the binding energy of SINM ensures that
the total energy obtained in the quite different variational spaces of
the HF and ETF methods are close, but does not guarantee that the individual 
contributions have the same size. We also recall that $a_{\text{surf}}$ is 
obtained as the difference between two large numbers that typically are two 
orders of magnitude larger, which tends to further amplify the differences 
between the methods.

The change of the contribution of the INM terms to $a_{\text{surf}}$
when going from SLy5s1 to SLy5s8 is reflected by the systematic changes of
INM parameters listed in Table~\ref{tab:parameters}. INM and SINM parameters
are clearly intertwined in a self-consistent model, and as long as the
parameters of $E_{\rho^{2}} + E_{\rho^{2 + \alpha}} + E_{\rho \tau}$ are
not kept fixed, the fit protocol shuffles contributions between them
in order to optimize the penalty function. Only about half of the change
of $a_{\text{surf}}$ when going from SLy5s1 to SLy5s8 originates in the
coupling constants of the gradient term $E_{\rho \Delta \rho}$. The
contribution of the spin-orbit term to $a_{\text{surf}}$ also varies slowly,
meaning that the present fit protocol interweaves the ``macroscopic''
and ``microscopic'' aspects of a parametrization: SLy5s1 with its smallest
$a_{\text{surf}}$ produces spin-orbit splittings that are about $10\%$
larger than those from SLy5s8, which has the largest $a_{\text{surf}}$
in the series. As is discussed below, this sometimes compromises the
possibility to distinguish the change in surface energy from changes in
shell effects when comparing parametrizations.

%
%==========================================================================
%
\section{Liquid-drop model estimates of surface energy}
\label{sec:LDM}

Below, we compare the results from self-consistent calculations with estimates
of total and deformation energies obtained from a LDM whose parameters are
extracted from the nuclear matter properties of the same effective
interaction.

For spherical nuclei, we use the following form of the LDM
energy, that is composed of volume, volume symmetry, surface and
surface symmetry energies as well as direct and exchange Coulomb
terms
\begin{eqnarray}
\label{eq:mac}
E_{\text{LDM}} (N,Z)
& = &   ( a_{\text{vol}}  + a_{\text{sym}} \, I^2 ) \, A
      \nonumber \\
&   &
      + ( a_{\text{surf}} + a_{\text{ssym}} \, I^2 ) \, A^{2/3}
      \nonumber \\
&   &
      + \frac{3 \, e^2}{5 \, r_0} \, \frac{Z^2}{A^{1/3}}
      - \frac{3 \, e^2}{4 \, r_0} \left( \frac{3}{2\pi}\right)^{2/3} \,
        \frac{Z^{4/3}}{A^{1/3}}
      \nonumber \\
&   &
\end{eqnarray}
where \mbox{$A = N+Z$} and \mbox{$I=\frac{N-Z}{N+Z}$}. The volume
($a_{\text{vol}}$) and volume symmetry ($a_{\text{sym}}$) energy
coefficients can be related to properties of INM at the
saturation point, whereas the surface ($a_{\text{surf}}$) and surface
symmetry ($a_{\text{ssym}}$) energy coefficients are connected to
properties of SINM. The radius constant $r_0$ entering
the Coulomb energies is determined by the nuclear matter saturation
density $\rho_{\text{sat}}$ through the relation
\mbox{$r_0^3 = 3/(4 \pi \rho_{\text{sat}})$}. The faithful
reproduction of binding energies from self-consistent calculations would
require additional higher-order terms \cite{Reinhard06a,Dobaczewski14a},
but this is irrelevant for the purpose of our further discussion.
We also omit the usual pairing term in Eq.~\eqref{eq:mac}, as its
parameters are mainly determined by the pairing functional~\eqref{eq:pair:edf}.

In general, the surface symmetry energy coefficient $a_{\text{ssym}}$
has the opposite sign of the surface energy coefficient $a_{\text{surf}}$,
which naturally follows from the volume and volume symmetry energy
coefficients also having the opposite sign.

For the following discussion it is useful to define an asymmetry-dependent
\textit{effective} surface energy coefficient
\begin{equation}
\label{eq:asurf:eff}
a_{\text{surf,eff}}(N,Z)
\equiv a_{\text{surf}} + a_{\text{ssym}} \, I^2 \, .
\end{equation}
As the value of the $a_{\text{ssym}}$ varies only very little among the
SLy5sX parametrizations, cf.\ Table~\ref{tab:parameters}, for a given
nucleus the difference between $a_{\text{surf,eff}}$ values of two
parametrizations remains very close to the difference between their
$a_{\text{surf}}$ values. For $^{240}$Pu, $a_{\text{surf,eff}}$ from
HF calculations of SINM takes the values of 15.29~MeV and 16.42~MeV for
SLy5s1 and SLy5s8, respectively. With 1.28~MeV, the difference between
these values differs only little from the 1.34~MeV difference between
the non-corrected values for $a_{\text{surf}}$. By contrast, for a
given parametrization $a_{\text{surf,eff}}$ can take visibly different
values for different nuclei: with 16.96~MeV and 18.24~MeV for SLy5s1
and SLy5s8, respectively, $a_{\text{surf,eff}}$ is 1.7~MeV larger for
the neutron-deficient $^{180}$Hg than for $^{240}$Pu. From this follows
immediately that the LDM fission barriers tend to decrease with asymmetry.

Mapping a self-consistent model on Eq.~\eqref{eq:mac} is, however,
non-trivial. The volume and volume symmetry energy coefficients,
$a_{\text{vol}}$ and $a_{\text{sym}}$, respectively, are directly given
by the INM properties $E/A$ and $J$ listed in Table~\ref{tab:INM}.
However, as already mentioned, determining $a_{\rm surf}$ of an
effective interaction has an inherent model-dependence.

The deformation dependence of the LDM energy~\eqref{eq:mac} is carried
by the Coulomb and surface energies. The latter can be parameterized
by multiplying the surface energy of the LDM formula~\eqref{eq:mac}
with a shape-dependent factor $B_s$~\cite{Hasse88} that is defined as
the ratio between the area of the surface of a deformed
liquid drop and a spherical one
\begin{equation}
\label{eq:Esurf:def}
E_{\text{LDM}}^{\text{surf}}(N,Z,\text{shape})
= \big( a_{\text{surf}} + a_{\text{ssym}} I^2 \big) \, A^{2/3} \,
  B_{s}(\text{shape}) \, .
\end{equation}
Because of volume conservation of the
nuclear liquid drop, this geometrical surface always grows with deformation.
The direct Coulomb energy of a deformed liquid drop with sharp surface
can in principle be parameterized through a similar factor
$E_{\text{LDM}}^{\text{C }}(N,Z,\text{shape})
= a_c \, Z^2 \, A^{-1/3} \, B_c(\text{shape})$ that has, however,
a different deformation dependence~\cite{Hasse88}. In the analysis of
deformation energy of finite nuclei in self-consistent calculations that
is presented below, we replace only the sum
$E_{\text{kin}} + E_{\text{Skyrme}} + E_{\text{corr}}$ by a LDM estimate,
while keeping $E_{\text{Coul}}$ from the self-consistent model.

For arbitrary parametrizations of the nuclear shape, the size of $B_{s}$
has in general to be determined through numerical integration of the surface
area. Similarly, the multipole moments of an arbitrarily
deformed liquid drop have in general also to be calculated through numerical
integration. For some specific shape parametrizations, however, both can be
developed in terms of a power series in the shape parameters~\cite{Hasse88}.
This can then be used to estimate the change of macroscopic
energy~\cite{Hasse88}
\begin{equation}
\label{eq:Edef}
E_{\text{def}}
=   E_{\text{surf}}
  - E_{\text{surf}}^{\text{sphere}}
  + E_{\text{Coul}}
  - E_{\text{Coul}}^{\text{sphere}} \, ,
\end{equation}
where the superscript ``sphere'' indicates the reference value of each term
for a spherical shape.

A widely-used parametrization of the nuclear surface for which such
analytical expressions exist is its expansion in spherical harmonics
\begin{equation}
\label{eq:dropexpansion}
R(\theta)
= \left[c(\alpha)\right]^{-1} \, R_0 \,
  \Big[ 1 + \sum_{\ell,m} \alpha_{\ell m} Y_{\ell m}(\theta) \Big] \, ,
\end{equation}
where $c(\alpha)$ is a normalization coefficient that ensures volume
conservation. Limiting ourselves to axially symmetric ($\alpha_{\ell m} = 0$
for $m \neq 0$) and reflection-symmetric ($\alpha_{\ell m} = 0$ for
odd $\ell$) shapes, relations given in Refs.~\cite{Myers83a,Hasse88}
can be used to express the ratio $B_s$ between the surface areas of a
deformed and a spherical liquid drop of the same volume~\eqref{eq:Esurf:def} as
\begin{eqnarray}
\label{eq:BsLDM}
B_s
& = & 1
      + \frac{1}{2\pi} \, \alpha_{20}^2
      - \frac{5}{210} \sqrt{\frac{5}{\pi}} \, \alpha_{20}^{3}
      - \frac{33}{556 \pi} \, \alpha_{20}^{4}
    \nn \\
&   &
      - \frac{3}{14 \pi \sqrt{\pi}} \, \alpha_{20}^2 \, \alpha_{40}
      + \frac{9}{4\pi} \, \alpha_{40}^2
\, .
\end{eqnarray}
A similar expression can be derived for the deformation dependence of the
Coulomb energy \cite{Poenaru11}.

In self-consistent models, however, the nuclear shape is naturally
characterized by multipole moments of the local (mass) density, which are
the expectation value of the operators
$\hat{Q}_{\ell m} \equiv r^\ell \, Y_{\ell m}(\vec{r})$. Their values
can be cast into the dimensionless deformations~\cite{Ryssens15a}
\begin{equation}
\label{equ:betalm}
\beta_{\ell m}
= \frac{4 \pi}{3 R_0^\ell A} \langle \hat{Q}_{\ell m} \rangle \, ,
\end{equation}
where $R_0 = 1.2 \, A^{1/3} \, \text{fm}$. These deformations are similar
in size to the shape expansion parameters of Eq.~\eqref{eq:dropexpansion},
but not equivalent. Indeed, adapting the expressions
of Refs.~\cite{Myers83a,Hasse88} to this expansion, for an axial liquid
drop characterized by $\alpha_{20}$ and $\alpha_{40}$, the corresponding
quadrupole and hexadecapole moments are
\begin{eqnarray}
\label{eq:droptransform}
\beta_{20}
& = & \alpha_{20}
      + \frac{2}{7}\sqrt{\frac{5}{\pi}} \, \alpha_{20}^2
      + \frac{20}{77} \sqrt{\frac{5}{\pi}} \, \alpha_{40}^2
      + \frac{12}{7\sqrt{\pi}} \, \alpha_{20} \, \alpha_{40}
    \nn \\
&   &
      - \frac{5}{28\pi} \, \alpha_{20}^3
      - \frac{235}{924\pi} \, \sqrt{\frac{5}{\pi}} \, \alpha_{20}^{4}
      + \frac{216 \sqrt{5}}{77\pi} \, \alpha_{20}^2 \, \alpha_{40} \, ,
    \nn \\
\beta_{40}
& = &   \alpha_{40}
      + \frac{9}{7\sqrt{\pi}} \, \alpha_{20}^2
      + \frac{300}{77\sqrt{5\pi}} \, \alpha_{20} \, \alpha_{40}
      \nn \\
&   &
      + \frac{275}{77\pi\sqrt{5}} \, \alpha_{20}^3
      + \frac{33975}{4004 \pi \sqrt{\pi}} \, \alpha_{20}^4 \, ,
\end{eqnarray}
where we have limited ourselves to fourth order in deformation, assuming that
$\alpha_{40}$ is on the order of $\alpha_{20}^2$. Note that the expressions
given in Ref.~\cite{Ryssens15a} use an inconsistent power counting, dropping
terms in $\alpha_{20}^3$ and $\alpha_{20}^{4}$ while keeping terms in
$\alpha_{40}^2$. The necessary extension to include also octupole distortions
is discussed for example in Refs.~\cite{Leander88,Wollersheim93}.

For a given self-consistent nuclear configuration characterized by
deformations $\{ \beta_{20}, \beta_{40}\}$,
Eq.~\eqref{eq:droptransform} can be numerically inverted to estimate
the expansion parameters $\alpha_{20}$ and $\alpha_{40}$
of a liquid drop that has the same multipole moments. From these values,
the corresponding LDM surface energy can then be estimated through
Eq.~\eqref{eq:Esurf:def}.

This approximate mapping of deformations can be expected to be reliable
only at small deformations for which powers of the $\alpha_{\ell 0}$ remain
smaller than $\alpha_{\ell 0}$ itself and where the higher-order deformations
that are neglected in Eq.~\eqref{eq:droptransform} do not play a significant
role yet.

For light nuclei, the LDM gives a spherical minimum and a broad single-humped
fission barrier that becomes lower with increasing charge number $Z$.
Neglecting the possible deformation dependence of the Coulomb exchange term,
for a LDM model parameterized through Eq.~\eqref{eq:Esurf:def} the fission
barrier vanishes when $E_{\text{def}} =
( a_{\text{surf}} + a_{\text{ssym}} I^2 ) A^{2/3} (B_s - 1)
+ a_c Z^2 A^{-1/3} (B_c - 1)$ becomes negative for arbitrary shape
distortions. Keeping only the leading term in the development of
$B_s(\text{shape}) \approx 1 + \tfrac{1}{2 \pi} \alpha_{20}^2 + \ldots$
from Eq.~\eqref{eq:BsLDM} and of
$B_c(\text{shape}) \approx 1 - \tfrac{1}{4 \pi} \alpha_{20}^2 + \ldots$
\cite{Poenaru11}
at small distortions of a sphere, this can be expressed through the
condition that the so-called \textit{fissility parameter}
$x$~\cite{Boh39a,Hasse88,Denschlag11}
\begin{equation}
\label{eq:fissility}
x
= \frac{E_{\text{Coul}}}{2 \, E_{\text{surf}}}
= \frac{3 \, e^2}{10 \, r_0}
  \frac{Z^2}
       {( a_{\text{surf}} + a_{\text{ssym}} \, I^2 ) \, A}
\end{equation}
becomes larger than 1. As a consequence,
very heavy nuclides with $x > 1$, which for typical values of the
LDM coefficients corresponds to $Z \gtrsim 104$, only exist because of quantal
shell effects. The exact location of the $x = 1$ line depends of course
on the values of the LDM parameters and can be very different for different
Skyrme EDFs~\cite{Nikolov11a}.

The concept of fissility has first been introduced for charged drops of
macroscopic liquids, for which it can also be directly experimentally
studied in great detail~\cite{Liao17}.

%===========================================================================

\section{Results}
\label{sec:results}

%---------------------------------------------------------------------------

\subsection{Numerical choices}

The calculations were all performed using the \texttt{MOCCa} code
\cite{RyssensPhd,MOCCa} that uses a coordinate-space representation.
It is based on the same principles as the published \texttt{EV8}
code~\cite{Bonche05,Ryssens15a}. The mesh parameter of
the Lagrange-mesh representation~\cite{Baye86a,Baye15a} is set to
$dx=0.8$~fm for all nuclei, with a suitable box size adapted to each case.
With the choices made, the numerical accuracy for the total
energies that are presented here is better than 100~keV, independent
on the deformation~\cite{Ryssens15b}.

It is well known that the correct description of the fission path usually
requires us to explore non-axial and octupole deformations.
The use of the MOCCa code permits us to do it in a general and consistent way.

Throughout this study, we use dimensionless mass multipole moments
\eqref{equ:betalm} to characterize deformation. Unless noted otherwise,
nuclei are oriented in such a way that axially symmetric states are
aligned with the $z$ axis. When plotting deformation energy curves as a
function of the quadrupole moment $\beta_{20}$, positive values of
$\beta_{20}$ indicate prolate shapes and negative values indicate oblate shapes.
We also discuss deformation energy surfaces of triaxial systems in the
$\beta$-$\gamma$ plane defined through~\cite{Ryssens15a}
\begin{eqnarray}
\label{eq:BM:beta}
\beta
& = & \sqrt{ \beta_{20}^2 + 2 \beta_{22}^2 } \, ,   \\
\label{eq:BM:gamma}
\gamma
& = & \text{atan2} \left( \sqrt{2} \,  \beta_{22} , \beta_{20} \right) \, .
\end{eqnarray}
To compare with experimental data for charge deformation obtained from
in-band $E2$ transitions, we also need charge deformations
$\beta_{\ell m,p}$, which are obtained from the multipole moments
$\langle \hat{Q}_{\ell m,p} \rangle$ of protons as
\begin{equation}
\label{equ:betalm:p}
\beta_{\ell m,p}
= \frac{4 \pi}{3 R_0^\ell Z} \langle \hat{Q}_{\ell m,p} \rangle \, .
\end{equation}
In a self-consistent mean-field model, values for $\beta_{\ell m,p}$
might differ on the percent level from the mass deformations
$\beta_{\ell m}$ defined through Eq.~\eqref{equ:betalm}.

%-----------------------------------------------------------------------------

\subsection{Fission barriers}
\label{sec:fission}

As fission barriers probe the deformation energy up to very large deformation
\cite{Bjornholm80,Denschlag11,Andreyev18r}, they are the natural starting
point to explore correlations between observables and the surface energy
coefficient. Here, we discuss three nuclei whose energy surfaces each have
a different topography.
\begin{itemize}
\item
The double-humped fission barrier of $^{240}$Pu has been used as the
reference case for many studies of different aspects of the fission process
and its modeling in an EDF framework 
\cite{Bartel82,Berger89,Rutz95,Bender03,Bender04a,Burvenich04a,Samyn05,Younes09,Li10,Abusara12,Schunck14,Jodon16}.

\item
The slightly lighter nucleus $^{226}$Ra is octupole-deformed in its
ground state and many calculations agree on the prediction of a
triple-humped fission barrier \cite{Rutz95}.

\item
The observation that the very neutron-deficient nucleus $^{180}$Hg fissions
asymetrically \cite{Andreyev10a} is an illustration that the shell effects
in highly-deformed configurations along the fission path determine
the most probable fission yields \cite{Andreyev10a,Ichikawa12,Veselsky12,Warda12,McDonnell14a},
not the shell effects in the fragments, which would favor a symmetric
split into two $^{90}$Zr.
\end{itemize}
This selection of systems has been motivated by the diversity of their
energy landscape, by their spread location in the chart of nuclei covering
a relatively wide range in mass and isospin, and by their fission
path that can be comparatively easily followed in calculations with a single
constraint. For many other systems, this is not the case, and
multidimensional calculations have to be carried out in order to reliably
find the saddle points \cite{Dubray12}.

Calculations have been performed assuming time-reversal invariance and
imposing two plane symmetries of the nuclear densities by choosing
the single-particle states to be eigenstates of $z$-signature $\hat{R}_z$
and the $y$-time-simplex $\hat{S}^T_y$.

Only the lowest continuous static fission path as obtained from calculations
allowing for both reflection-asymmetry and non-axiality is presented.
We first discuss the changes in the topography of the energy curves of these
three nuclei when systematically varying the surface energy coefficient,
and compare with experiment later on.

%- - - - - - - - - - - - - - - - - - - - - - - - - - - - - - - - - - - - - -
%
\subsubsection{Fission barrier of $^{240}$Pu}

\begin{figure}[t!]
\includegraphics[width=8.5cm]{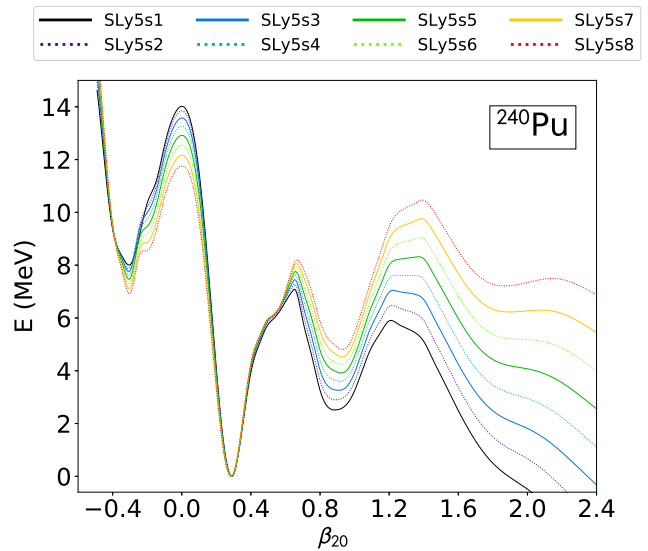}
\caption{\label{fig:Pufission}
Deformation energy curve of $^{240}$Pu as a function of
$\beta_{20}$ for the parametrizations as indicated.
The energies are normalized to the respective ground-state energy.
}
\end{figure}

The fission barrier of $^{240}$Pu is shown in Fig.~\ref{fig:Pufission}.
We have checked that the configurations along the fission path change
continuously without sudden jumps. The fission path is practically the
same for all SLy5sX parametrizations, with the deformation parameters
$\beta_{\ell m}$ \eqref{equ:betalm} taking near-identical values.

Up to the superdeformed minimum associated with the fission isomer, the
lowest configurations are reflection symmetric. For larger deformations,
octupole deformation sets in. Around the top of the inner and outer barriers
at $\beta_{20} \simeq 0.5$ and $\beta_{20} \simeq 1.3$, respectively, the
saddle points are lowered by non-axial shapes, by about 1.5~MeV for the
inner barrier and about 0.5~MeV for the outer barrier.\footnote{When
calculating the energy curves of $^{240}$Pu as obtained
with the SLy5sX reported in Ref.~\cite{Jodon16} we failed to find the
non-axial solution of the outer barrier. The energy curves shown there
differ from Fig.~\ref{fig:Pufission} also by the use of HF+BCS instead
of HFB.}
The corresponding $\gamma$ angles \eqref{eq:BM:gamma} go up to about
12~degrees for the inner and 1.5~degrees for the outer barrier, which
corresponds to values of $\beta_{22}$ of about 0.07 and 0.02, respectively.

With $\beta_{20,p}$ taking values between 0.293 for SLy5s1 and 0.287 for
SLy5s8, the ground-state deformation agrees very well with the charge quadrupole
deformation $\beta_{20,p} = 0.293 \pm 0.002$ that can be deduced with
the usual expressions \cite{Bender04a} from the experimental
$B(E2;0^+ \to 2^+)$ value \cite{Bemis73}. Similarly, the calculated
values for the charge hexadecapole deformation $\beta_{40,p}$ that fall
between 0.164 for SLy5s1 and 0.155 for SLy5s8 also agree with the value
$\beta_{40,p} = 0.166 \pm 0.040$ extracted from the measured
$B(E4;0^+ \to 4^+)$ that has been reported in Ref.~\cite{Bemis73}.
Note that in the full $\beta$-$\gamma$ plane, the oblate saddle at
$\beta_{20} \simeq -0.3$ is connected without barrier to the prolate
minimum through triaxial shapes.

As expected, going from SLy5s8 with its high value of $a_{\text{sym}}$ to
SLy5s1 with its low one, the deformation energy relative to the
ground state is significantly reduced for states that have a larger
deformation than the ground state. This reduction is quite uniform, such
that at a given deformation the curves are almost equally spaced, and their
spread almost uniformly increases with deformation.
That the order of the curves is inverted around the spherical point and for
oblate shapes is a consequence of normalizing all energies relative
to the deformed ground state. Qualitatively, the same spread of the energy
curves that is visible when going from the prolate ground state to the
superdeformed state also happens when going from the
spherical state to the deformed ground state, although on a smaller energy
scale. But since energies are normalized to the more deformed one among the
two states, the order of the curves becomes inverted when going from the
ground state to smaller deformations instead of larger ones.
The same artifact from normalization also appears on several other
plots discussed in what follows.

The increasing change of deformation energy as a function of deformation
that accompanies a change in $a_{\text{surf}}$ can make shallow minima appear
or disappear, as can be seen from the third hyperdeformed minimum at
$\beta_{20} \simeq 1.8$ that is predicted by SLy5s8. Going to
parametrizations with lower $a_{\text{surf}}$ it gradually vanishes as
the downfalling slope of the energy becomes increasingly steep.

Quantitatively, reducing $a_{\text{surf}}$ by $7.2 \%$ when going from SLy5s8
to SLy5s1 reduces the outer barrier by about $5 \, \text{MeV}$, which is
about $40 \%$ in this case. The relative height of the inner and outer
barriers is also reversed: for SLy5s8 it is the outer barrier that is the
highest one, whereas for SLy5s1 it is the inner barrier. Both differences
would have an enormous impact on fission dynamics calculated with one or
the other of these parametrizations.

%
%- - - - - - - - - - - - - - - - - - - - - - - - - - - - - - - - - - - - - -
%
\subsubsection{Fission barrier of $^{226}$Ra}

\begin{figure}[t!]
\includegraphics[width=8.5cm]{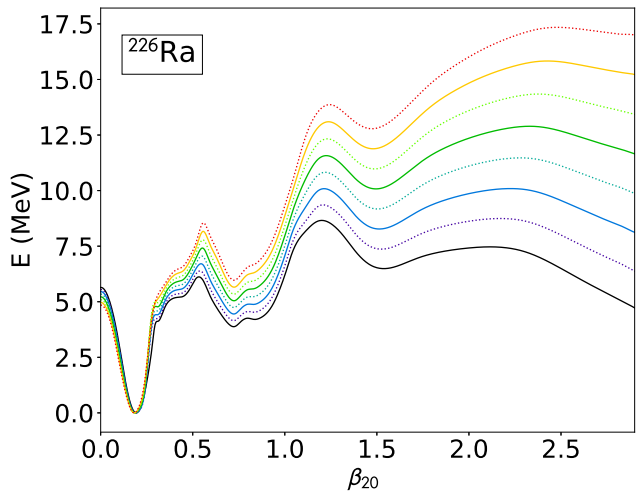}
\caption{\label{fig:Rafission}
Same as Fig.~\ref{fig:Pufission}, but for $^{226}$Ra.
}
\end{figure}

The energy curves for $^{226}$Ra are shown in Fig.~\ref{fig:Rafission}.
Again, all parametrizations give the same fission path.
This nucleus is located in a small region of the nuclear chart where
stable octupole deformations are present in the ground state,
leading to an energy gain of a couple of 100~keV compared to the
reflection-symmetric saddle. The values are slightly increasing with
decreasing $a_{\text{surf}}$, from about 250~keV for SLy5s8
to about 700~keV for SLy5s1. The actual octupole deformation of the
ground state also sensitively depends on the parametrization, and
increases with decreasing $a_{\text{surf}}$ from $\beta_{30} = 0.06$
for SLy5s8 to $\beta_{30} = 0.12$ for SLy5s1. A similar behavior is
found for the majority of nuclides with octupole-deformed ground state,
see the discussion in Sect.~\ref{sec:octupole}.
Around $\beta_{20} \simeq 0.5$, the first
barrier proceeds through a region of reflection-symmetric, but triaxial
configurations. Around the superdeformed minimum at $\beta_{20} \simeq 0.7$,
the lowest states become  axial and reflection-symmetric. From the second
barrier onwards, the lowest states take reflection-asymmetric shapes,
including the ones in the third minimum around $\beta_{20} \simeq 1.5$.
A detailed comparison of the lowest energy curve obtained with SLy5s1
with the ones obtained when imposing reflection symmetry and/or axiality
can be found in Ref.~\cite{Ryssens16}. We have not found any reduction of
the outer barriers when allowing for non-axial shapes.

The calculated ground-state proton quadrupole moment takes values
between $\beta_{20,p} = 0.161$ (SLy5s8) and 0.169 (SLy5s1), which
somewhat underestimates the empirical value $\beta_{20,p} = 0.202(3)$
determined from the experimental $B(E2;0^+ \to 2^+)$ value
\cite{Wollersheim93} for all parametrizations.

The evolution of differences between parametrizations with deformation
are qualitatively the same as what was found for $^{240}$Pu. As the
outermost barrier is at larger deformation, its overall reduction when
going from SLy5s8 to SLy5s1 is even more dramatic. Again the difference of
barrier heights changes sign, here for the second and third barriers. The
broad third barrier found with SLy5s8 becomes almost becomes a shoulder
with SLy5s1, thereby making the third minimum very shallow.

%- - - - - - - - - - - - - - - - - - - - - - - - - - - - - - - - - - - - - -
%
\subsubsection{Fission barrier of $^{180}$Hg}

\begin{figure}[t!]
\includegraphics[width=8.5cm]{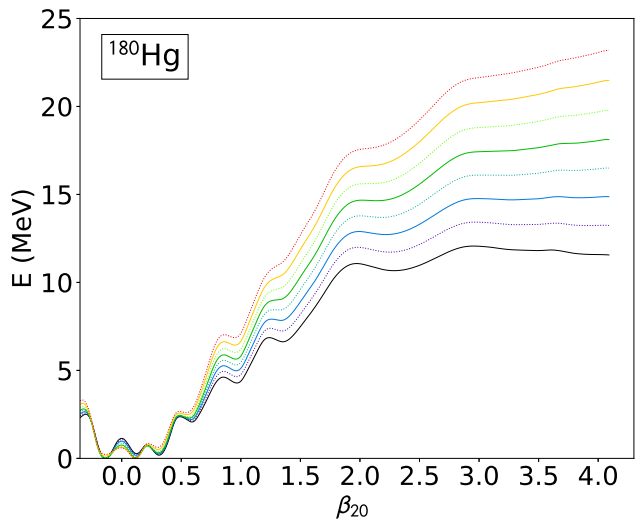}
\caption{\label{fig:Hgfission}
Same as Fig.~\ref{fig:Pufission}, but for $^{180}$Hg.
}
\end{figure}

Figure~\ref{fig:Hgfission} displays the fission barrier of $^{180}$Hg.
Again, all parametrizations lead to the same fission path. The energy
landscape of this nucleus differs from the other two discussed above in
several respects: First, it exhibits shape coexistence at small
deformation. Second, there is only one broad barrier, whose saddle point
is at much larger deformation.
Third, the scission point, where the fissioning nucleus breaks apart, is
very close to the top of the barrier. The curves in Fig.~\ref{fig:Hgfission}
end where the calculations jump to a different solution with two
non-identical fragments that is about 30~MeV below.
There are also several superdeformed and hyperdeformed local minima
visible between the normal-deformed minima and the barrier, a feature already
found in earlier calculations of this nucleus. We have not checked
if the barriers separating these structures become lower or even disappear
when allowing for non-axial shapes. The broad outer barrier follows a
reflection-asymmetric path beginning at around $\beta_{20} \simeq 1.1$,
such that the shallow minimum at slightly larger deformation exhibited
by SLy5s1 corresponds to octupole deformed shapes. We have not found
non-axial solutions that lower the barrier at these large deformations.

The evolution of differences between parametrizations with deformation
is again qualitatively the same as what was found above for $^{240}$Pu and
$^{226}$Ra. As the outermost barrier is at larger deformation, the
overall variation of barrier height is even more dramatic. Again, the
position of the saddle point changes: it moves from $\beta_{20} \simeq 4.0$
to $\beta_{20} \simeq 3.0$ when going from SLy5s8 to SLy5s1.

We mention in passing that the deformation of the ground state changes
from weakly oblate for parametrizations with low $a_{\text{surf}}$ up to
SLy5s4 to weakly prolate for parametrizations with higher $a_{\text{surf}}$.
The same also happens for some of the adjacent Hg isotopes, which has an
impact on the evolution of charge radii and their odd-even staggering.
As has been discussed elsewhere \cite{Sels18x},
the parametrizations with low $a_{\text{surf}}$ provide a much better
description of these data than the ones with high $a_{\text{surf}}$.

%- - - - - - - - - - - - - - - - - - - - - - - - - - - - - - - - - - - - - -

\subsubsection{Correlation between characteristic energies and $a_{\text{surf}}$}
\label{sect:correlation}

\begin{figure}[t!]
\includegraphics[width=8.5cm]{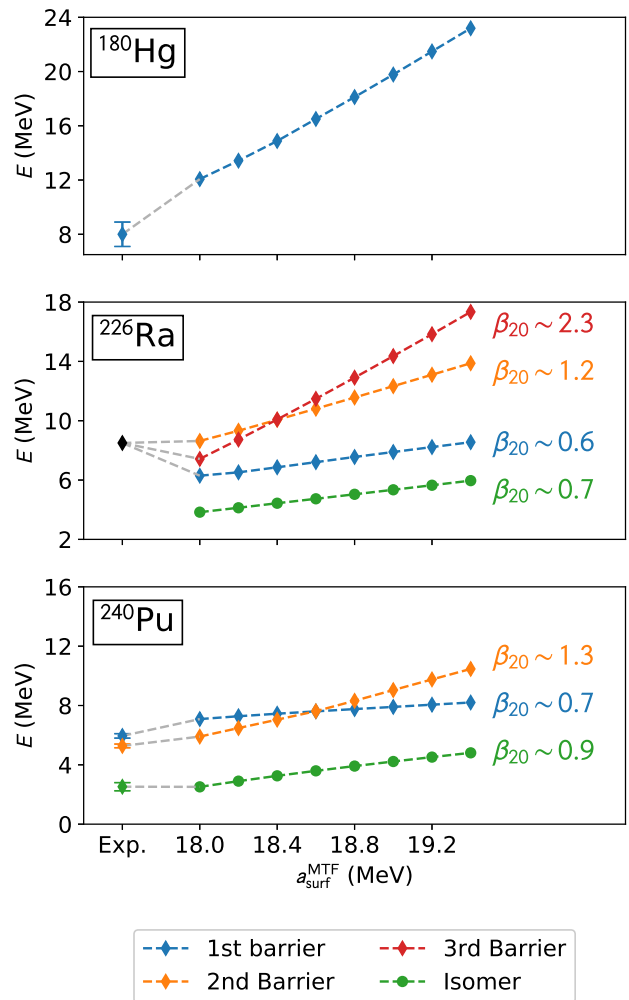}
\caption{\label{fig:compact}
Height of the first ($E_{\rm 1st}$), second ($E_{\rm 2nd}$) and third
($E_{\rm 3rd}$) barrier as well as the excitation energy of the fission isomer
($E_{\rm iso}$) of the nuclei as indicated versus the surface energy
coefficient $a_{\rm surf}$ calculated in the MTF model as used as a
constraint in the fit of the SLy5sX parametrizations.
}
\end{figure}

The energy curves presented above indicate that for the
SLy5sX parametrizations the differences between deformation energies
scale with the surface energy coefficient $a_{\text{surf}}$ and
the deformation. This is confirmed when plotting some characteristic
energies such as the heights of various barriers and excitation energies
of the superdeformed states for these nuclei directly as a function of
$a_{\text{surf}}$, see Fig.~\ref{fig:compact}.

The deformation dependence \eqref{eq:Esurf:def} of the LDM \eqref{eq:mac}
predicts that for a given nucleus at a given shape with fixed deformation
parameters the total energy is a linear function of $a_{\text{surf}}$.
And indeed, all curves are almost linear, with the most pronounced
deviation for the barrier height of $^{180}$Hg, although even there the
slight bend of this curve is just barely visible. The reason for the latter
is that the saddle point gradually changes its deformation from
$\beta_{20} \simeq 4$ to $\beta_{20} \simeq 3$ when lowering
$a_{\text{surf}}$, cf.\ Fig.~\ref{fig:Hgfission}.

This plot confirms the validity of the idea of Ref.~\cite{Jodon16} that the
adjustment of interactions to characteristic energies of fission barriers
can be replaced by the adjustment of a suitably chosen value of
$a_{\text{surf}}$.

To the left in Fig.~\ref{fig:compact}, we provide experimental data, where
available, for comparison.
We recall, however, that the empirical determination of fission barrier
heights and excitation energies of fission isomers is not trivial and, in
general, requires the application of some model. Calculated fission barriers
are usually those of the ground state, whereas in experiment they are most
often probed through the decay of excited states. For $^{240}$Pu, the error
bars cover the range of the values reported
in Refs.~\cite{Capote09,Mamdouh98,Hunyadi01,Samyn05,Bjornholm80}.
The barrier height of 8.5~MeV for $^{226}$Ra is the recommended value
from the RIPL-3 database~\cite{Capote09}. The situation is more complicated
for $^{180}$Hg, for which data come from the observation of $\beta$-delayed
fission of $^{180}$Tl, which passes through excited states of negative
parity and finite angular momentum at an excitation energy that is
necessarily smaller than the $Q$ value for electron capture of $^{180}$Tl,
$Q_{\text{EC}}({}^{180}\text{Tl}) = 10.44 \, \text{MeV}$. The model-dependent
analysis of the measured probability of $\beta$-delayed fission in that
nucleus \cite{Veselsky12} suggests that the fission barrier has a height of about
$8.0(9) \, \text{MeV}$, which is the value used in Fig.~\eqref{fig:compact}.

None of the parametrizations reproduces all data simultaneously, but it is
obvious that those with the smallest values of $a_{\text{surf}}$ in the set
are clearly favored. Compared to the other parametrizations, SLy5s1 gives
a fair description of the available data for $^{240}$Pu and $^{226}$Ra, but
still significantly overestimates the barrier for $^{180}$Hg. The latter,
however, is the most uncertain value in the data set. Macroscopic-microscopic
models that describe well the fission barriers of actinide nuclei also
give a barrier height of $^{180}$Hg that is too high by about 2 MeV
\cite{Veselsky12}. We recall that many widely used parametrizations of the
Skyrme EDF such as SLy4 have a surface energy coefficient that, when
calculated in the MTF model, take values about $19.0 \pm 0.3$ MeV \cite{Jodon16}
and then overestimate fission barriers, while the often-used standard
for fission studies, SkM*, has an $a_{\text{surf}}$ of a similar size as
SLy5s1.

There are many possible reasons for the scatter of the deviations from
data. First, shell effects might be incorrect, either in the ground state
or for the saddle point or both. Indeed, the complicate structure of the
deformation energy curves of Figs.~\ref{fig:Pufission}, \ref{fig:Rafission},
and \ref{fig:Hgfission} with its multiple maxima and minima at various
deformations is determined by shell effects. Without them, these three
nuclei would have a spherical ground state and one broad fission barrier
without any substructure.
Second, the isospin dependence of the surface energy of the
SLy5sX might be incorrect. This, however, cannot be corrected for by simply
increasing the absolute value of the (negative) surface symmetry energy
coefficient $a_{\text{ssym}}$ in Eq.~\eqref{eq:mac}. With
$I^2 = (20/180)^2 \approx 0.012$, the asymmetry of $^{180}$Hg is much smaller
than the one of $^{240}$Pu, $I^2 = (52/240)^2 \approx 0.047$ such that the
reduction of the barrier of the latter would be greater. Instead, one has
to change both $a_{\text{surf}}$ and $a_{\text{ssym}}$ in order to reproduce
both barriers simultaneously.
Assuming that the entire change of the outer fission barrier heights of
$^{180}$Hg and $^{240}$Pu between SLy5s8 and SLy5s1 is due to the
change in their effective surface energy coefficient~\eqref{eq:asurf:eff},
one can estimate values that would describe both barriers through linear
extrapolation. Reproducing the outer barrier height of 5.3~MeV of $^{240}$Pu
with a parametrization adjusted along the lines of the SLy5sX
calls for an effective surface energy coefficient

$a_{\text{surf,eff}}^{\text{HF}}(146,94) \simeq 15.14 \; \text{MeV}$,
whereas reproducing the recommended value of 8.0~MeV for the barrier height
of $^{180}$Hg from Ref.~\cite{Veselsky12} demands for
$a_{\text{surf,eff}}^{\text{HF}}(100,80) \simeq 16.49 \; \text{MeV}$. To
obtain such values requires us to set
$a_{\text{surf}}^{\text{HF}} \approx  16.97 \; \text{MeV}$ and
$a_{\text{ssym}}^{\text{HF}} \approx -39 \; \text{MeV}$.
Using the upper limit of 8.9~MeV for the estimated barrier height
of $^{180}$Hg instead yields
$a_{\text{surf}}^{\text{HF}} \approx 17.11 \; \text{MeV}$ and
$a_{\text{ssym}}^{\text{HF}} \approx -42 \; \text{MeV}$
instead. With that, the ratio of both would be reduced from
$a_{\text{ssym}}^{\text{HF}}/a_{\text{surf}}^{\text{HF}} \approx -2.7$ for all
SLy5sX to -2.3, and become closer to the ratio of the volume coefficients,
which takes values of about $a_{\text{vol}}/a_{\text{sym}} \approx 2$
for all SLy5sX. This analysis, however, assumes that shell effects
are correctly described in both nuclei, which is not necessarily the case.
While pushing the values for $a_{\text{surf}}^{\text{HF}}$ and
$a_{\text{ssym}}^{\text{HF}}$ slightly outside the range of combinations
found for the majority of parametrizations of Skyrme EDF, the changes
suggested by the above analysis remain comparatively small, such that
there is no \textit{a priori} reason to rule them out. In fact, the
modified values estimated from the barrier heights
of $^{240}$Pu and $^{180}$Hg become very close to those
of the modern Lublin-Strasbourg drop (LSD) parametrization of the LDM
\cite{Pomorski03},
$a_{\text{surf}}^{\text{LSD}} =  16.9707 \; \text{MeV}$ and
$a_{\text{ssym}}^{\text{LSD}} = -38.9274 \; \text{MeV}$,
which has been adjusted to masses and fission barrier heights. Because of the
model-dependence of the actual values of the surface energy coefficients
extracted from EDFs \cite{Jodon16,Meyer18x}, however, it cannot be ruled out
that this agreement is fortuitous.

\begin{figure}[t!]
\includegraphics[width=8.0cm]{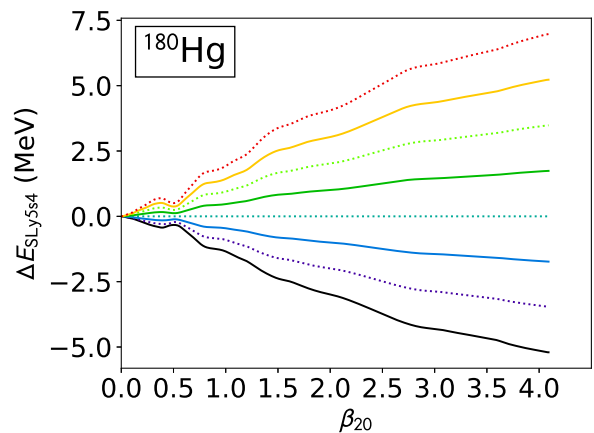}
\caption{\label{fig:Hg180:diff}
Difference $E^{\text{SLy5sX}}(\text{shape}) - E^{\text{SLy5s4}}(\text{shape})$
between the deformation energy of $^{180}$Hg as obtained with
the SLy5sX parametrizations and shown in Fig.~\ref{fig:Hgfission} and the
deformation energy obtained with SLy5s4.
The same color code as in Fig.~\ref{fig:Pufission} is used.
}
\end{figure}

We recall that the simple near-linear dependence of
$E_{\text{def}}(\text{shape})$ as a function of $a_{\text{surf}}$
as exhibited by Fig.~\ref{fig:compact} can only be found when using a series
of parametrizations that have been adjusted within the same dedicated
protocol. Otherwise there is a large scatter around the global linear
trends as exemplified in Ref.~\cite{Jodon16}.

\begin{figure}[b!]
\includegraphics[width=8.0cm]{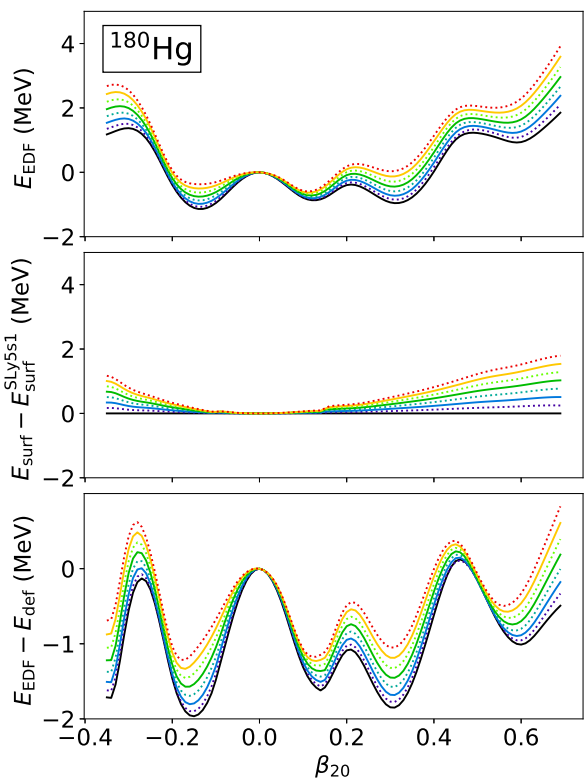}
\caption{\label{fig:HgLDM}
Deformation energy of $^{180}$Hg as obtained from self-consistent
calculations (top), difference between the surface energies
$E_{\text{LDM}}^{\text{surf}}(\text{shape})$, defined
through~\eqref{eq:Esurf:def},
of a deformed liquid drop~\eqref{eq:dropexpansion} that yields the
same multipole deformations $\beta_{20}$ and $\beta_{40}$ as the
self-consistent calculation (middle), and the evolution of shell
effects estimated by subtracting the LDM deformation energy~\eqref{eq:Edef}
from the energy of the self-consistent calculations.
All curves are drawn as a function of the axial quadrupole deformation
$\beta_{20}$ and normalized to the energy at spherical shape, and
the same color code as in Fig.~\ref{fig:Pufission} is used to distinguish
the SLy5sX parametrizations. Note that the same energy interval is used
for the upper two panels.
}
\end{figure}

There is the question how much, and for which configurations, shell effects
do actually differ when going from one parametrization to another.
Figure~\ref{fig:Hg180:diff} shows the difference in deformation energy
between SLy5s4 and the other SLy5sX parametrizations for prolate
configurations up to the scission point. The differences are clearly not
simple straight lines as would be the case if the SLy5sX parametrizations
only differed in their surface energy coefficient. Instead, there are
oscillations around the linear trend that occur at the same deformation for
all parametrizations, and whose amplitude increases when going to smaller
deformation. The amplitude of these oscillations also increases with the
difference in $a_{\text{surf}}$ of the parametrizations. For this reason
we choose to plot the difference to an intermediate parametrization, which
allows for a better resolution of the deviations
than plotting the difference to either SLy5s1
or SLy5s8. The deformations where the most pronounced of these oscillations
are situated correspond to the barrier below the SD minimum
($\beta_{20} \approx 0.4$), the SD minimum ($\beta_{20} \approx 0.6$),
and the barriers between the highly-deformed minima
($\beta_{20} \approx  0.4$, 0.9, 1.2, \ldots).

Limiting the discussion to small deformations below $\beta_{20} = 0.7$ where
the expansion~\eqref{eq:dropexpansion} of the nuclear surface can be expected
to approximately match the actual shape of the self-consistent
solutions, Fig.~\ref{fig:HgLDM} provides estimates of the contribution
from the LDM surface energy and shell effects to the deformation energy
of $^{180}$Hg.

The upper panel shows again the deformation energy as obtained from
self-consistent calculations, but unlike Fig.~\ref{fig:Hgfission} the
energies are now normalized to the spherical shape. To estimate the LDM
surface energy, the deformations $\alpha_{20}$ and $\alpha_{40}$ entering
the expression for the deformed liquid drop~\eqref{eq:dropexpansion} are
determined from the deformations $\beta_{20}$ and $\beta_{40}$ of the
self-consistent states by inversion of relations~\eqref{eq:droptransform}.
From these, the deformation-dependent LDM surface energy
$E_{\text{LDM}}^{\text{surf}}(\text{shape})$ is calculated by
inserting~\eqref{eq:BsLDM} into~\eqref{eq:Esurf:def}. This is done separately
for the results obtained with each SLy5sX parametrization, using the
respective effective surface energy coefficient~\eqref{eq:asurf:eff}
constructed from the HF values of $a_{\text{surf}}$ and $a_{\text{ssym}}$
taken from Table~\ref{tab:parameters}.
As we are only interested in relative changes, the LDM surface energy obtained
from SLy5s1 is subtracted from all curves. The actual LDM deformation energy
changes by about 25~MeV over the range of the plot, most of which is
compensated by an almost as large change in Coulomb energy. The shapes
found along the path of lowest energy are almost independent on the
parametrization, such that the difference between the LDM surface energies
$E_{\text{LDM}}^{\text{surf}}(\text{shape})$ of the various SLy5sX is simply
monotonically growing with $\beta_{20}$ as it is entirely determined
by the differences in $a_{\text{surf,eff}}$. However, as the surface area
of a liquid drop, Eq.~\eqref{eq:Esurf:def}, is a complicated function
of $\beta_{20}$ with the leading term being
quadratic, the curves for the $E_{\text{LDM}}^{\text{surf}}(\text{shape})$
are not straight lines themselves. It is only at a given deformation that the
distance between two curves can be expected to be proportional to the
difference in $a_{\text{surf}}$ between the parametrizations they were
constructed with.

Having deduced the LDM deformation energy, we can also estimate
``microscopic'' shell energies that are at the origin of the minima and
maxima of the self-consistent energy curve. Such quantity can be constructed
by subtracting the deformation dependent parts of the LDM surface energy
$E_{\text{LDM}}^{\text{surf}}(\text{shape})$ and
the self-consistent Coulomb energy, calculated through Eq.~\eqref{eq:Edef},
from the total binding energy from the EDF. The resulting energy curves,
normalized to the spherical point, are shown in the lower panel.
For obvious reasons their local minima coincide with the local minima of
the energy curves from self-consistent calculations in the upper panel,
and the maxima with the barriers in between.

Comparing the results obtained with different SLy5sX parametrizations,
the amplitude of the variations of shell effects along the path of lowest
energy differs on the scale of
about 1~MeV. It is smallest for SLy5s8 and largest for SLy5s1.
This variation is correlated to the size of the spin-orbit coupling constant
$C^{\rho \nabla \cdot J}_{0}$, see Table~\ref{tab:parameters}, which has a
significant impact on the position of some specific single-particle states
near the Fermi surface, see the discussion of Fig.~\ref{fig:Hg:Nilsson}
in what follows.

It has to be stressed that this ``microscopic'' energy is not completely
equivalent to a shell correction as defined through the Strutinski theorem
\cite{Nilsson95a,Brack72a,Brack75a,Brack81a}. Besides imperfections of
the mapping on the LDM surface energy, this energy also contains the
neglected higher-order contribution to the LDM energy, the
deformation-dependence of the pairing energy, as well as the deformation
dependence of the tensor terms $E_{JJ}$ in Eq.~\eqref{eq:skyrme:energy},
which depends on the filling of shells. In addition, only the variation
of this microscopic energy with deformation can be determined from our
mapping, not its absolute size.

Still, the results indicate that, in spite of their identical fit protocol,
rearrangement effects during the parameter fit lead to changes between
the SLy5sX parametrizations such that the shell effects of the ground state
of $^{180}$Hg vary by about 1~MeV. For other nuclei the changes might
even be larger, as the amplitude of the variation of shell effects itself
is larger. Unfortunately, the analysis of shell effects outlined above
becomes much more involved when non-axial or reflection-asymmetric
deformations come into play, as is the case for the static fission paths
of $^{226}$Ra and $^{240}$Pu.

%===========================================================================
%
\subsection{Superheavy nuclei}
\label{sect:SHE}

\begin{figure}
\includegraphics[width=8.0cm]{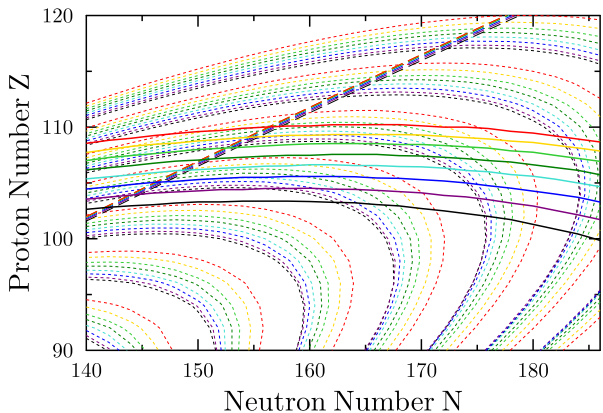}
\caption{\label{fig:SHE:LDM}
Gross properties of superheavy nuclei as obtained from the LDM parameters
of the SLy5sX parametrizations. Using the same color code for the
parametrizations as in Fig.~\ref{fig:Pufission},
the light dotted lines show the contours of $E/A$ of spherical nuclei, varying
between $-7.6$~MeV in the lower left corner and $-7.0$~MeV in the upper left
corner in steps of 100~keV. The heavy dashed lines show the proton drip line
defined through $S_{2p}(N,Z) = 0$, whereas the heavy solid lines indicate
the line where fissility takes the value $x = 1$.
}
\end{figure}

In the discussion of \nuc{180}{Hg}, \nuc{226}{Ra} and \nuc{240}{Pu},
we have seen that the outer saddle point tends to move
to smaller deformation. As outlined in Section~\ref{sec:LDM}, the LDM
fission barrier becomes smaller with increasing charge number $Z$ until
it vanishes for nuclei for which the fissility $x$, as defined in
Eq.~\eqref{eq:fissility}, becomes larger than one. In fact, fissility is
an often used criterion to define the so-called superheavy elements as
those that have a vanishing liquid-drop fission barrier and only exist because
the quickly fluctuating shell effects still generate a fission barrier
\cite{Ackermann17}.

Figure~\ref{fig:SHE:LDM} shows the line of fissility $x=1$ in the region of
known transactinide nuclei, calculated by inserting the INM parameters as
listed in Table~\ref{tab:INM} and the HF values for the surface and surface
symmetry energy coefficients from Table~\ref{tab:parameters} in
Eq.~\eqref{eq:fissility}. The figure also shows the lines of equal
$E_{\text{LDM}}/A$ and the position of the two-proton dripline, where
$S_{2p}(N,Z) \equiv E_{\text{LDM}}(N,Z-2) - E_{\text{LDM}}(N,Z)$ becomes
zero.

The relative displacement of lines of equal $E/A$ indicates clearly that
for the SLy5sX parametrizations the (negative) LDM binding energy
$E_{\text{LDM}}(N,Z)$ of a given nucleus takes larger absolute values when
increasing $a_{\text{surf}}$.
This clearly points to large rearrangement effects during the parameter
fit, as increasing $a_{\text{surf}}$ in the LDM energy~\eqref{eq:mac}
while keeping all other parameters constant would decrease the absolute
value of the
binding energy. By contrast, the position of the two-proton drip-line
calculated from the LDM binding energies is practically the same. This
is the consequence of another rearrangement effect in the coupling
constants during the fit that is elaborated in more detail in
Sect.~\ref{subsect:isotopic} below.

We note in passing that, because of its $A$ and $I^2$ dependence, the value
of the fissility is not the same for all isotopes of an element with given
$Z$, which introduces some ambiguity into the definition of superheavy
elements as those for which $x > 1$.

\begin{figure}[t!]
\centerline{\includegraphics[width=4.2cm]{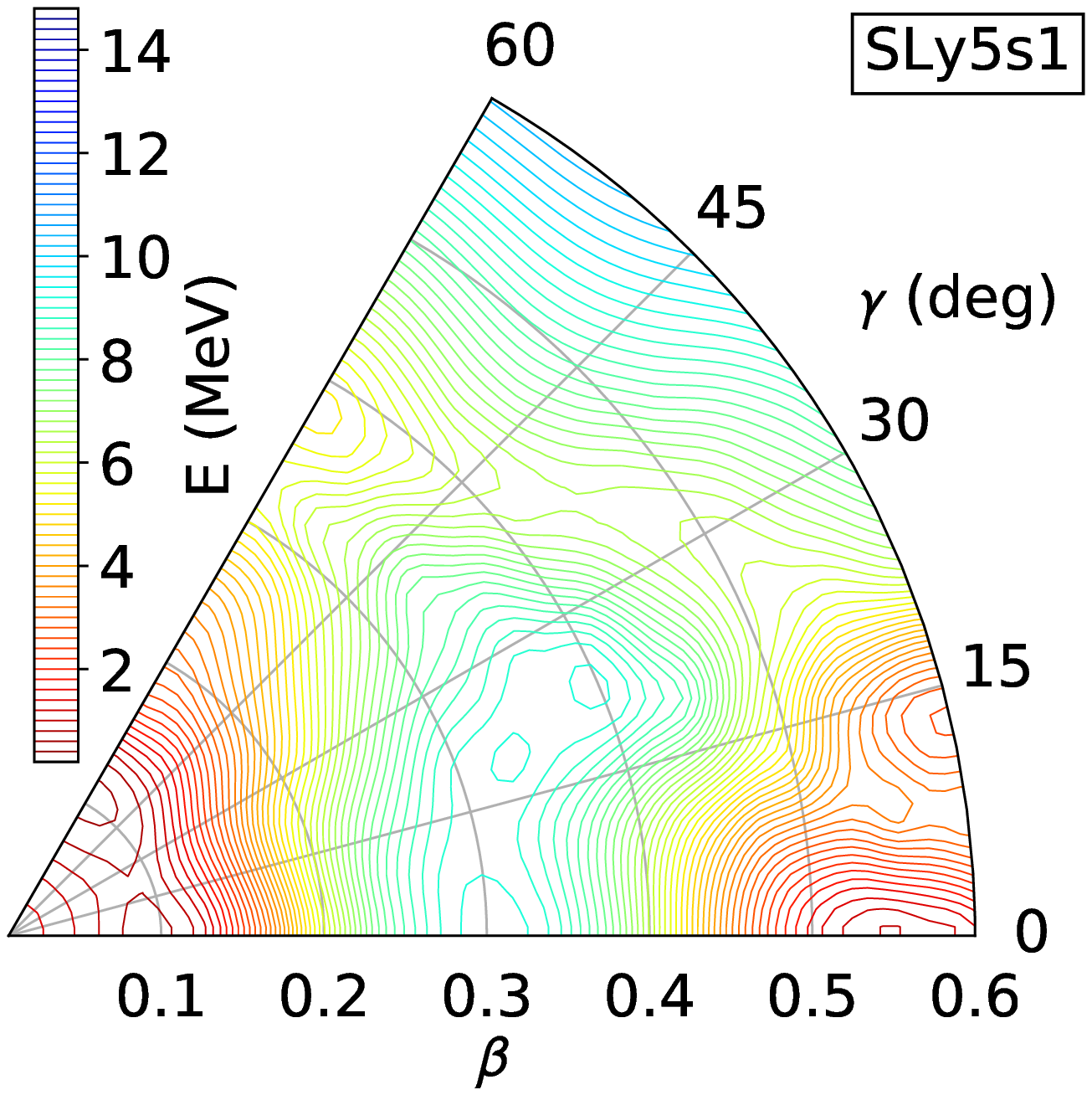}
            \includegraphics[width=4.2cm]{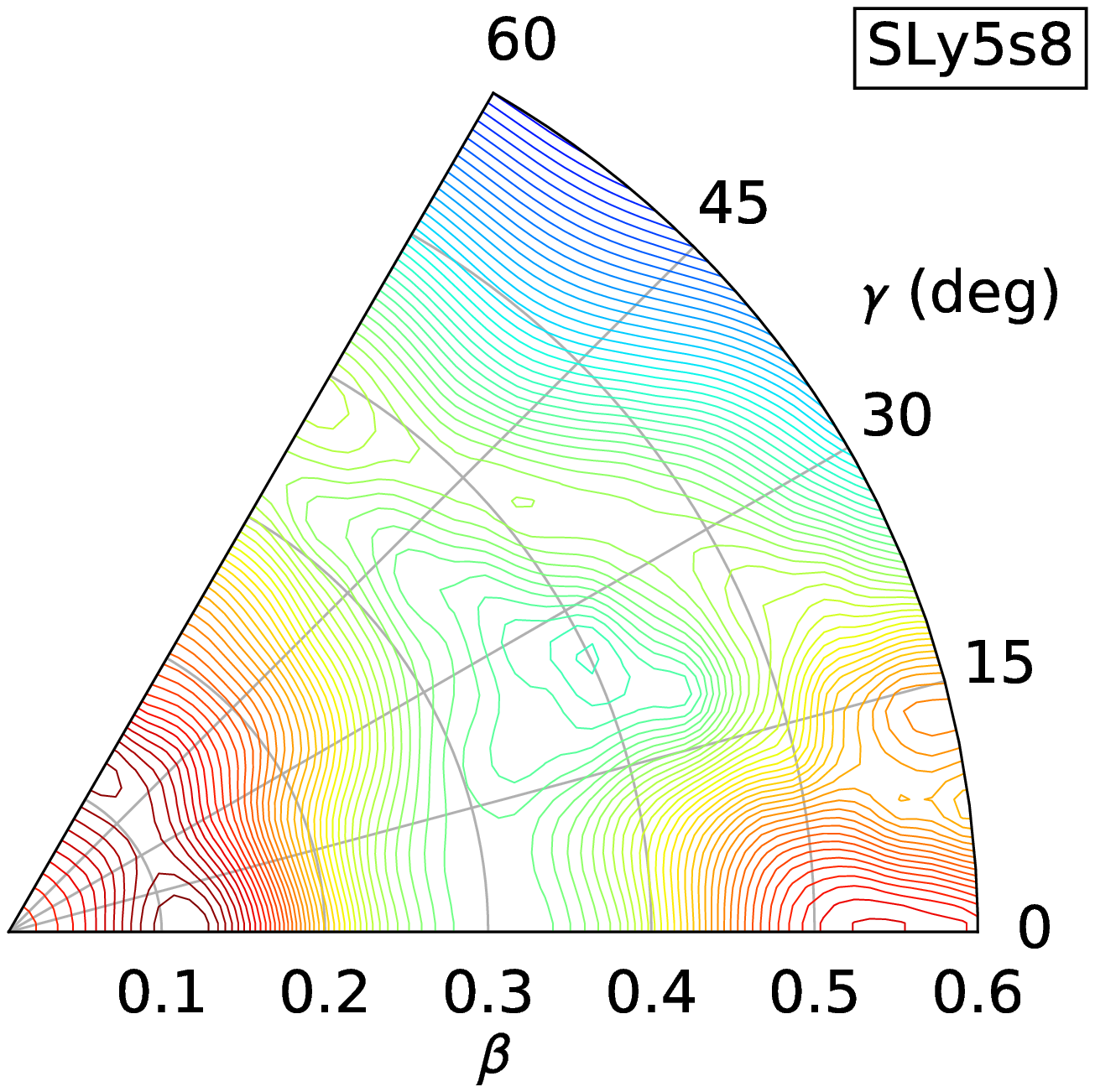}}
\caption{\label{fig:Og}
Deformation energy surface of $^{294}$Og using the SLy5s1 (left) and
SLy5s8 (right) parametrizations from calculations assuming
reflection-symmetric triaxal shapes. The energies are normalized to the
respective ground state minimum that has been determined by interpolation
of the calculated points on the surface.
}
\end{figure}

As an example of the deformation energy of a superheavy nucleus,
Fig.~\ref{fig:Og} displays the energy surfaces in the $\beta$-$\gamma$
plane of $^{294}$Og ($Z=118$, $N=176$), that is the heaviest
even-even nucleus identified in experiment so far \cite{294118}. Results
are only shown for SLy5s1 and SLy5s8, the parametrizations at the extremes
of the interval of $a_{\text{surf}}$. Unlike the cases of lighter nuclei
discussed above, the fission path is not exactly the same, although its
gross features remain similar for both parametrizations.

This nucleus, which is in a region of transitional nuclei close to the
next spherical shell closures, has a quite complicated deformation
energy surface, such that a path of lowest energy cannot be easily
calculated and represented as a function of a single deformation
parameter. Indeed, at deformations below $\beta_2 \lesssim 0.15$,
the energy surface is quite flat, with the absolute minimum being on the
prolate side for SLy5s8, whereas it is oblate for SLy5s1. The static fission
path proceeds through axial oblate shapes before turning towards triaxial
shapes. It bypasses an axial superdeformed prolate minimum at
$\beta_{20} \approx 0.55$ and proceeds through a very shallow
excited triaxial minimum at $\beta_{2} \approx 0.6$,
$\gamma \approx 15^\circ$ instead. Beyond the border of the plot,
the energy of that valley continues to fall off. The outer barrier
is further lowered when also allowing for non-axial reflection-asymmetric
shapes, but we have not systematically explored this degree of freedom.
In any event, the inner barrier is higher than the outer one.

The difference in barrier height between SLy5s1 and SLy5s8 on
this path of lowest energy remains very small; for the former it is
about 6.1 MeV with the saddle reached for oblate shapes, while for the
latter it is about 6.5 MeV reached at triaxial shapes.
This is much smaller than what has been found for the differences between
inner barriers of $^{226}$Ra or $^{240}$Pu.
The reasons for this different behavior are
that on the one hand the underlying LDM energy surface is very flat,
and that on the other hand the variation of shell effects is clearly
not the same as evidenced by the many small differences in the topography
of the energy surfaces. This is particularly obvious when looking at the
excited fission path that passes through near-axial shapes on the prolate
side of the $\beta$-$\gamma$ plane: with slightly more than 8.4~MeV,
it is higher for SLy5s1 than for SLy5s8, which
gives only about 7.6~MeV. It is only at larger deformation beyond
$\beta_{2} \gtrsim 0.5$ that the energy surface obtained with SLy5s8
is visibly stiffer than the one from SLy5s1 as one would naively expect.
The axial superdeformed minimum has about 0.25~MeV excitation energy for
SLy5s1, whereas it is at 1.25~MeV for SLy5s8.

\begin{figure}[t!]
\centerline{\includegraphics[width=4.2cm]{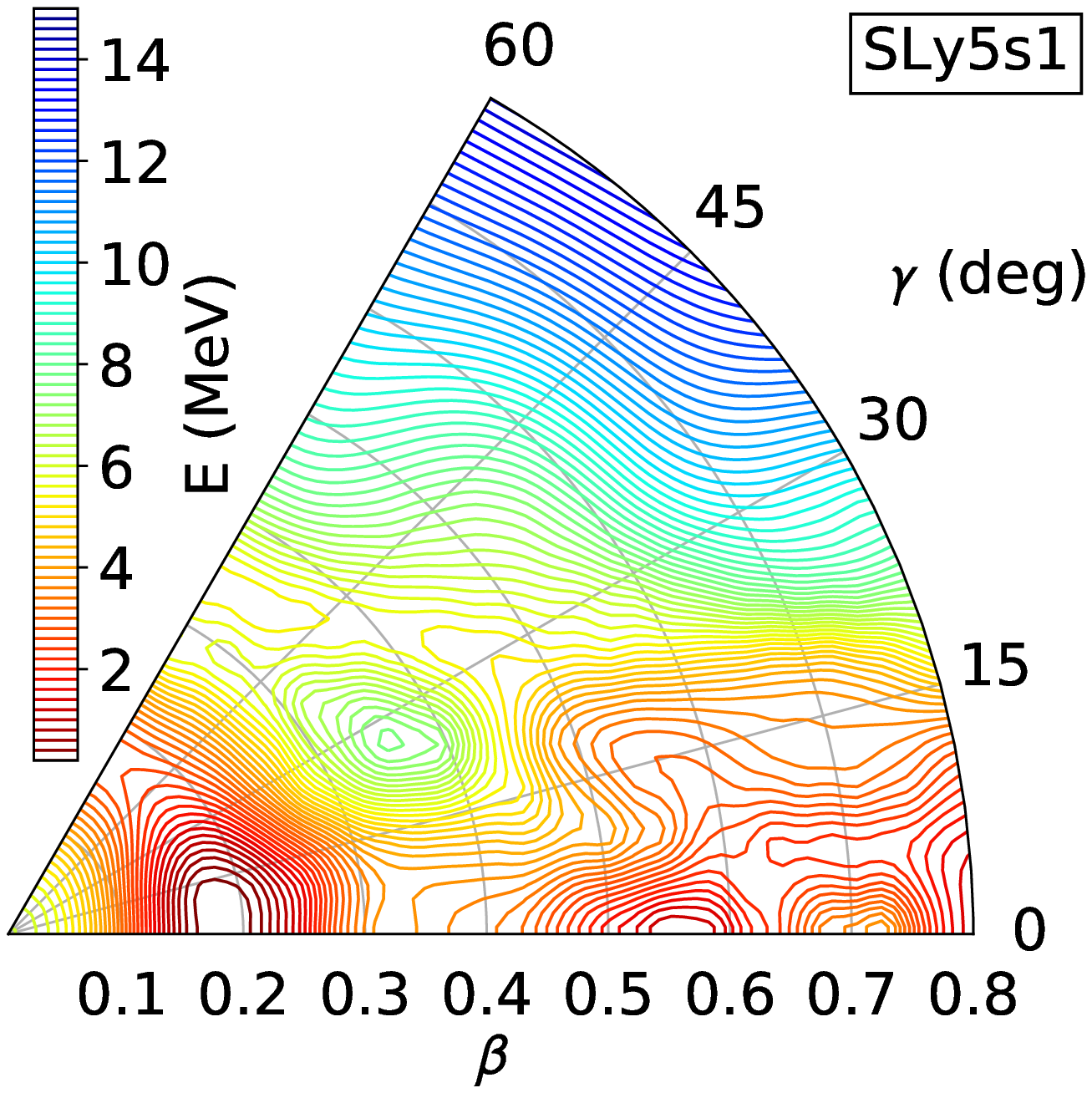}
            \includegraphics[width=4.2cm]{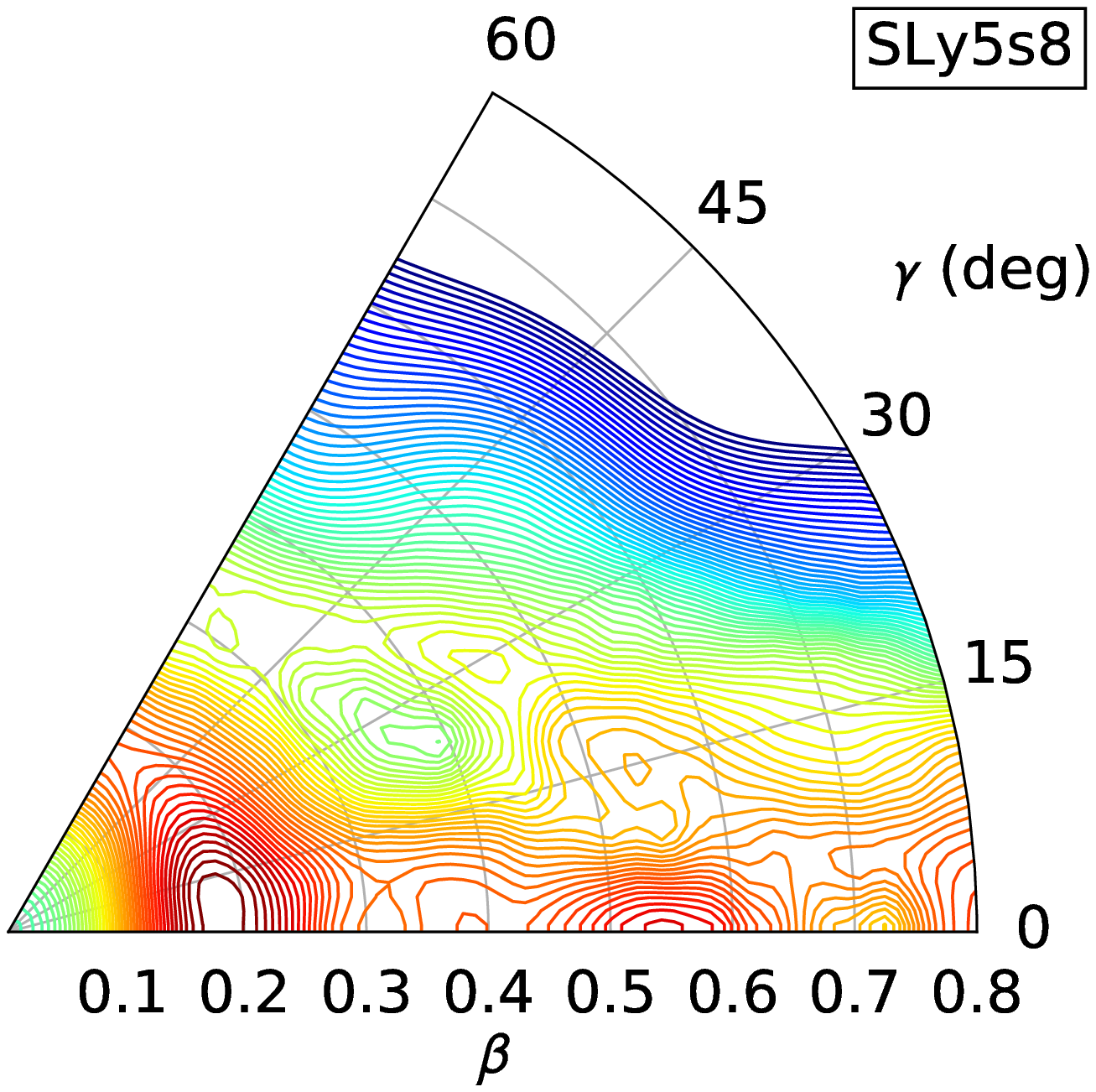}}
\caption{\label{fig:Cn}
Same as Fig.~\ref{fig:Og}, but for $^{282}$Cn.
}
\end{figure}

The few observed $\alpha$-decay chains of $^{294}$Og end with
$^{282}$Cn ($Z=112$, $N=170$) \cite{294118}, a nucleus whose decay
is dominated by spontaneous fission. The energy surfaces obtained for
this nucleus are shown in Fig.~\ref{fig:Cn}. Passing through near-axial
prolate shapes, the calculated static fission path of $^{282}$Cn is rather
similar to the one of heavy actinides. Both parametrizations agree on a
prolate ground state, and also give a shallow excited axial superdeformed
minimum at $\beta_{20} \simeq 0.55$, whose excitation energy
is less than 1~MeV. With about 3.6~MeV, the inner barrier is higher
for SLy5s1 than for SLy5s8, for which it takes a value of about 3.0~MeV,
similar to what has been found for the excited near-axial path for
$^{294}$Og. It is again only for the outer barrier that SLy5s8 gives
a larger height than SLy5s1, as one would expect from their values for
$a_{\text{surf}}$. For both parametrizations, the outer barrier passes again
through triaxial shapes, and continues to fall off outside the border
of the plots. With 3.2~MeV, it is minimally higher than the inner one for
SLy5s8, whereas for SLy5s1 it is the other way around. Reflection-asymmetric
shapes, not considered when preparing Fig.~\ref{fig:Cn}, lower again the
energy surface for $\beta_{2} \gtrsim 0.55$, such that it is the inner
barrier which determines this nucleus' lifetime.

There also is a shallow excited valley connecting an excited triaxial
minimum with oblate shapes through triaxial ones with large $\gamma$
values, similar to the static fission path found for $^{294}$Og in
Fig.~\ref{fig:Og}. On this path, with 6.2~MeV the saddle from SLy5s8
is higher than the one at 5.6~MeV predicted by SLy5s1, similar to
what has been found for $^{294}$Og.

In any event, the fission barrier of $^{282}$Cn is correctly predicted
to be much lower than the one of $^{294}$Og. However, these two examples
illustrate that for superheavy nuclei at the limits of the presently known
chart of nuclei there is not necessarily a direct correlation between
fission barrier heights and the value of $a_{\text{surf}}$.
The structure of these nuclei with fissility larger than one is dominated
by shell effects that, in spite of the common fit protocol, turn out to
be slightly different for the SLy5sX parametrizations.

%=========================================================================

\subsection{Superdeformed minima}
\label{sect:SD:min}

\begin{figure}[t!]
\includegraphics[width=8.0cm]{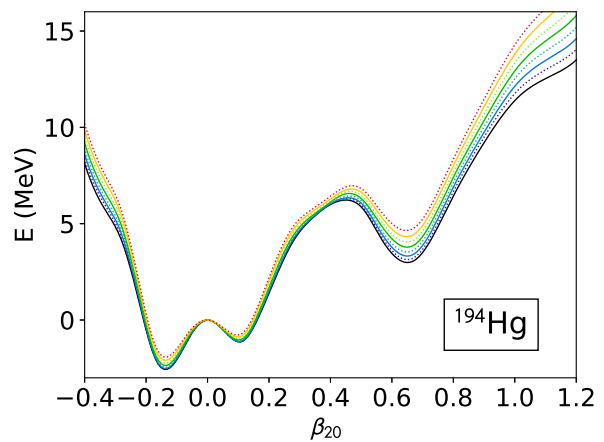}
\caption{\label{fig:Hg194}
Energy of $^{194}$Hg as a function of axial quadrupole
deformation parameter $\beta_{20}$ for the SLy5sX parametrizations using
the same color code as in Fig.~\ref{fig:Pufission}. The energy is normalized
to the spherical configuration.
}
\end{figure}

Throughout the chart of nuclei, there are regions where one can find excited
rotational bands with very large moments of inertia that extend to
high spins well beyond $I \gtrsim 40$. These bands can be associated to shapes
that have much larger deformations than what is found for ground states,
with $\beta_{20}$ taking values between 0.6 and 0.8. In heavy nuclei,
such deformation brings
single-particle levels originating from two major spherical shells above
or below close to the Fermi energy. Because of the resulting
significant difference in occupied orbits, electromagnetic decay out of
these bands to normal-deformed (ND) states is in general hindered. Thanks
to that, the bands can be followed over many transitions down
to an end point where the decay out of the band takes place abruptly,
with a very complicated highly-fragmented multi-step pattern of transitions.
In many cases, the decay-out cannot be resolved in experiment \cite{Lopez16}.
This phenomenon has been dubbed superdeformation (SD) in the literature.

The SD bands found for nuclei in the neutron-deficient $A \simeq 190$ region
are of special interest for our analysis, because for some of them the
decay-out to ND yrast states occurs at low spins of about 10 $\hbar$,
and with only a few intermediate steps that can be resolved in experiment.
From the known excitation energies of states in the SD band, the
excitation energy of the (unobserved) SD $0^+$ band-head can be estimated
with a good accuracy
\cite{Khoo96,Hauschild97,Lauritsen00,Wilson03,Wilson05,Wilson10}.

\begin{figure}[t!]
\includegraphics[width=8.0cm]{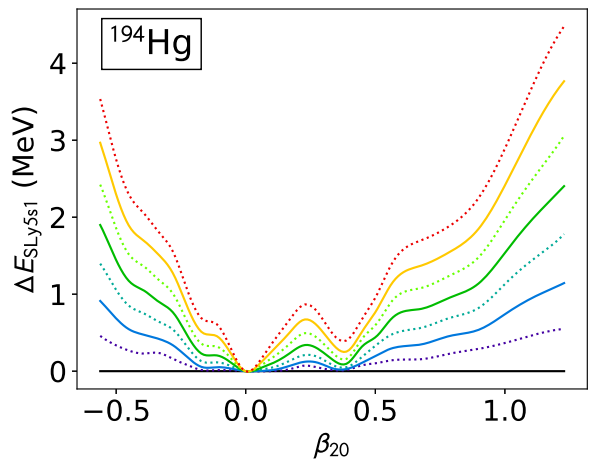}
\caption{\label{fig:Hg194:diff}
Difference between the deformation energy of $^{194}$Hg obtained with the
SLy5sX parametrizations relative to the one obtained with SLy5s1. The same
color code as in Fig.~\ref{fig:Pufission} is used.
}
\end{figure}

One of these nuclides is $^{194}$Hg. Figure~\ref{fig:Hg194} shows the
variation of the total energy of this nucleus as a function of the
axial quadrupole deformation. Besides normal-deformed prolate and oblate
minima there is a superdeformed one at $\beta_{20} \simeq 0.65$.
As for the other nuclei discussed so far, the overall
topography of the energy surface is the same for all eight parametrizations,
and differences between them grow larger as the deformation increases. The
most pronounced minima are obtained with SLy5s1. With increasing
$a_{\text{surf}}$, the energy surface becomes overall stiffer, part of
which is, however, because of a significant difference in shell effects
between the parametrizations, similar to what has been found for $^{180}$Hg.
This becomes clearly visible
when plotting the difference between the deformation energy obtained from
different parametrizations; see Fig.~\ref{fig:Hg194:diff}. The curves do
indeed not rise monotonically as they would if the surface tension
were the only source of differences between the deformation energies.

\begin{figure}[t!]
\includegraphics[width=8.0cm]{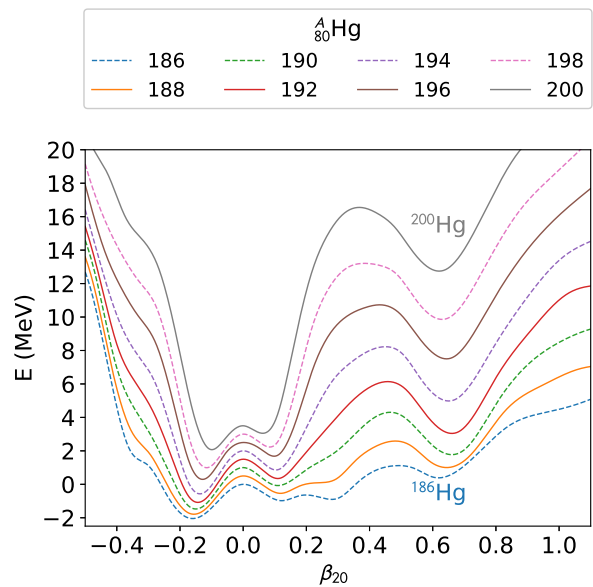}
\caption{\label{fig:Hg:all:def:SLy5s1}
Deformation energy of even-even Hg isotopes as a function of axial quadrupole
moment $\beta_{20}$ for the SLy5s1 parametrization.
The energy curves are drawn with an offset of 0.5~MeV for the spherical
state between two consecutive nuclei.
}
\end{figure}

Other neutron-deficient Hg isotopes in the $A \approx 190$ region exhibit
energy curves with very similar structure, although the relative energy
between the various minima and the height of the barriers separating them
changes when going down in neutron number from $N=120$ to $N=106$,
see Fig.~\ref{fig:Hg:all:def:SLy5s1}. From the rapid variation of the
excitation energy of the SD minimum and the barrier that separates it
from ND states it is clear that there is a strong variation of neutron
shell effects at large deformation with neutron number. The deformation
of the SD minimum varies only very little, however, which indicates that
it is mainly caused by a proton shell effect.

\begin{figure}
\label{fig:nilssonHg}
\includegraphics[width=8.8cm]{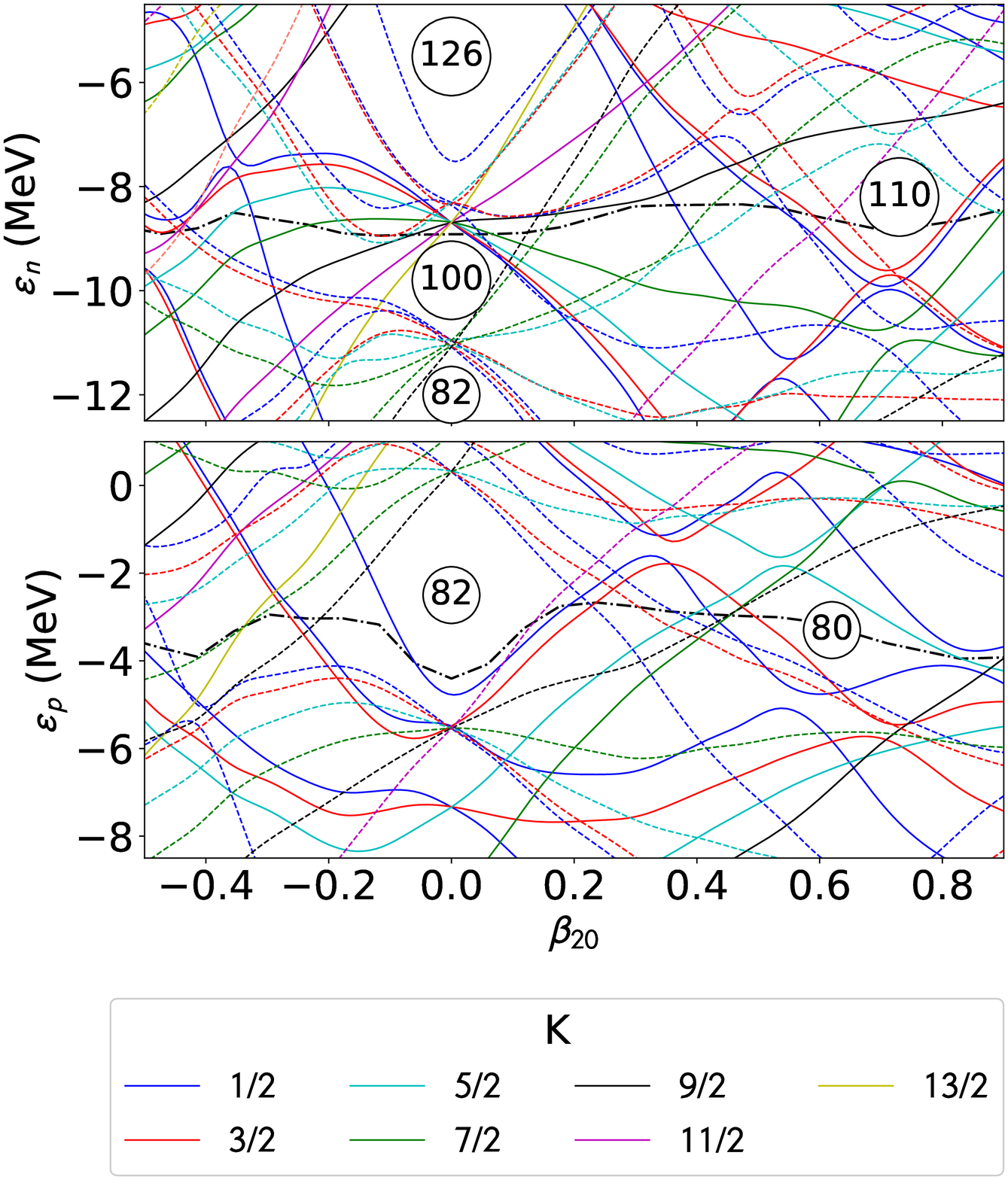}
\includegraphics[width=8.8cm]{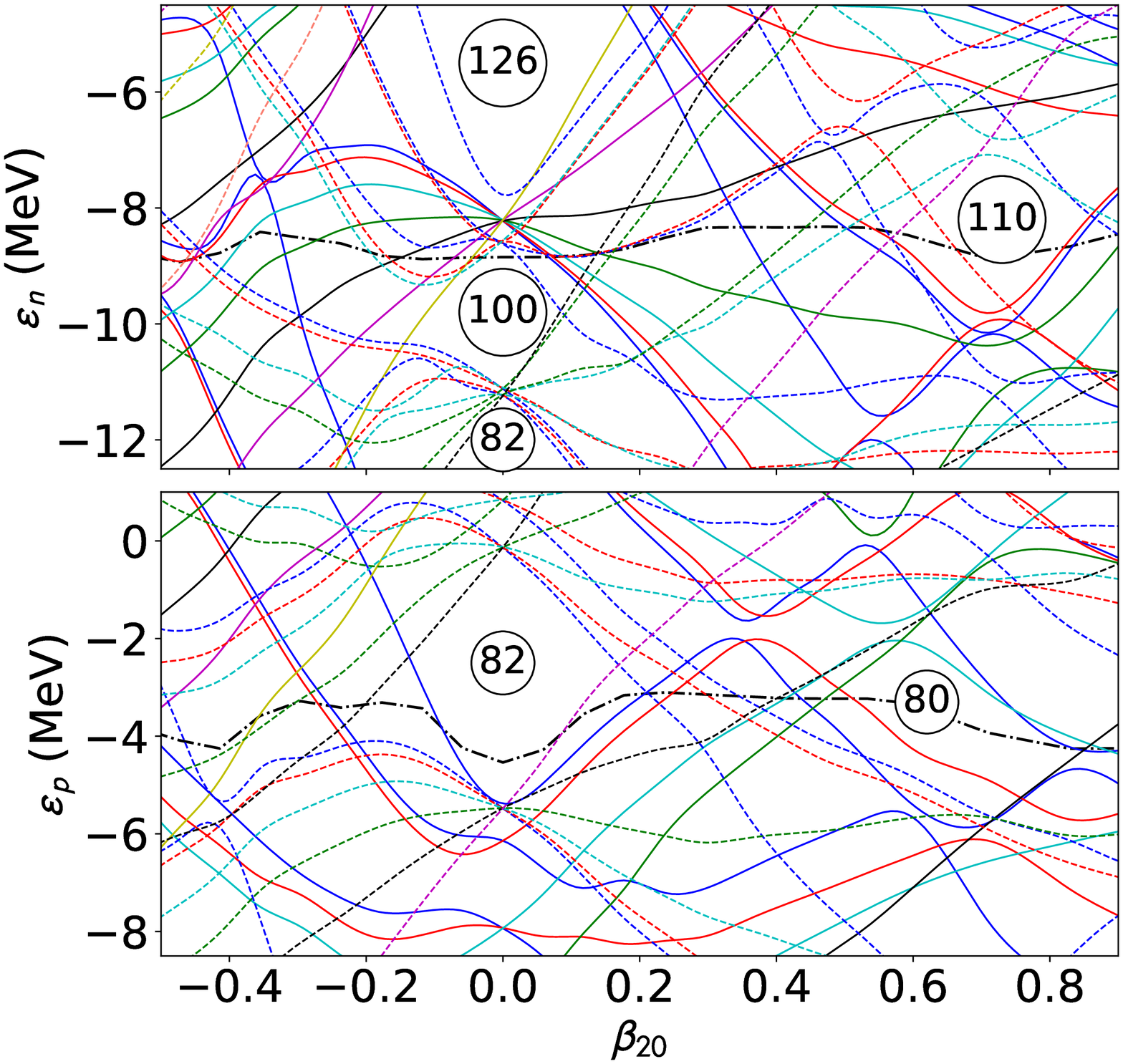}
\caption{\label{fig:Hg:Nilsson}
Nilsson diagrams of eigenvalues of the single-particle Hamiltonian
of neutrons ($\epsilon_n$) and protons ($\epsilon_p$) for axially
symmetric states of $^{188}$Hg obtained with the SLy5s8 (top) and SLy5s1
parametrizations. The color of the lines
indicates the expectation value of $\langle \hat{j}_z\rangle$ for the
respective single-particle state: solid (dashed) lines correspond to levels
of positive (negative) parity. The respective Fermi energies of protons and
neutrons are drawn as dash-dotted black lines.}
\end{figure}

This is confirmed by the Nilsson diagrams of these nuclei.
Figure~\ref{fig:Hg:Nilsson} shows examples for $^{188}$Hg, obtained with
SLy5s1 and SLy5s8. There is a deformed $Z=80$ gap at the deformation
$\beta_{20} \simeq 0.65$ of the SD minimum. There also is a large deformed gap
for neutrons at $N=110$, but at slightly larger deformation
$\beta_{20} \simeq 0.75$, which is not sufficient to generate an additional
minimum. Instead, its main effect seems to be to soften or stiffen the energy
curve at deformations larger than the one of the SD minimum. We
recall that the shell effects on energy curves are generated by large
deviations of the level density around the Fermi energy from the average
one, and not by the actual size of the gaps between the levels. As the
bunching of neutron levels above and below the $N=110$ gap is much larger
than what is found for proton levels around the $Z=80$ gap, the net neutron
shell effect on the energy curves is much weaker than the one of the
protons, in spite of the size of the actual gap being larger.

At the deformation of the SD minimum, there are smaller neutron gaps
for $N=112$ and $N=114$ that, however, are not of the same size for SLy5s1
and SLy5s8. This can be related to small differences between the relative
position of spherical shells that can be attributed to the difference of
spin-orbit coupling constants $C^{\rho \nabla \cdot J}_{t}$; cf.\
Table~\ref{tab:parameters}. Spin-orbit splittings tend to be slightly larger
for SLy5s1 than for SLy5s8, which then has a visible influence on the
deformation dependence of shell effects that can be seen in
Fig.~\ref{fig:Hg194:diff}. Its most obvious effect at spherical shape
concerns the position of the high-$j$ intruder level near the Fermi
surface relative to the low-$j$ levels around it. For SLy5s8 with its
comparatively small $C^{\rho \nabla \cdot J}_{t}$ the neutron $i_{13/2^+}$
intruder level is between the $p_{1/2^-}$ and the quasi-degenerate $f_{5/2^-}$
and $p_{3/2^-}$ levels, whereas for SLy5s1, with its $10 \, \%$ larger values
of $C^{\rho \nabla \cdot J}_{t}$, the $i_{13/2^+}$ is pushed below the latter
two levels. Similarly, for protons the $h_{11/2^-}$ intruder level is
quasi-degenerate with the $s_{1/2^+}$ level, whereas for SLy5s1 it is almost
1~MeV lower. These changes have a significant impact on the shell effects
at small deformation for all nuclei in this region of the chart of nuclei,
as exemplified already by Fig.~\ref{fig:Hg194:diff}.

The excitation of the SD minimum of $^{194}$Hg has been studied earlier with
traditional parametrizations of the Skyrme EDF in
Refs.~\cite{Takahara98,Nikolov11a}. The range of values found is larger than
the one covered by the SLy5sX parametrizations.

We mention in passing that the appearance of the prolate ND minimum
at $\beta_{20} \approx 0.3$ is associated with a ND prolate deformed
band observed for isotopes with $N \approx 104$. Its excitation energy
is notoriously difficult to describe \cite{Yao13,Bree14a,Wrozek18x}.
In this respect, a very delicate observable
is the odd-even staggering of charge radii of Hg isotopes between $^{180}$Hg
and $^{186}$Hg, for which the prolate minimum remains an excited state for
even-even nuclei, but becomes the ground state for the odd-$A$ isotopes in
between \cite{Sels18x}. Among the parametrizations discussed here, this
behavior is best, although not perfectly, described by the parametrizations
with low $a_{\text{surf}}$ up to SLy5s4 \cite{Sels18x}.

\begin{figure}
\includegraphics[width=8.5cm]{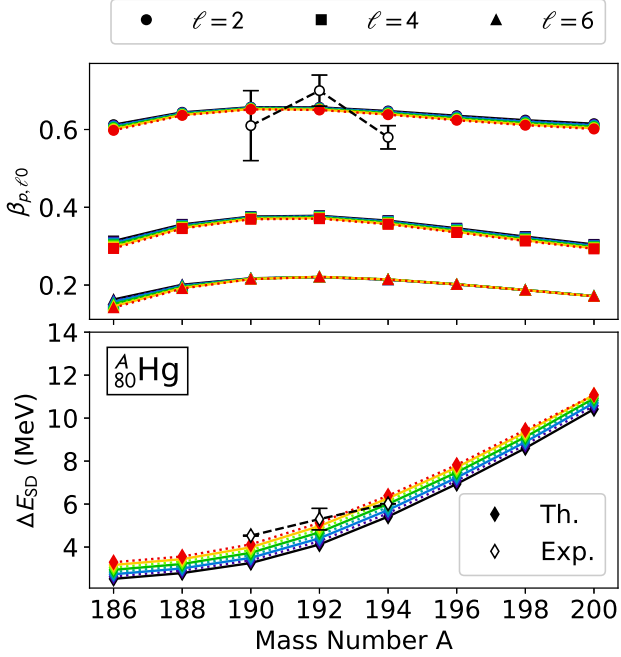}
\caption{\label{fig:Hgsuper}
Excitation energy $\Delta E_{\text{SD}}$with respect to the
ground state and multipole deformations $\beta_{\ell 0}$ of the proton
distribution of the superdeformed minimum for the even-even Hg isotopes
as indicated and compared with experimental data where available (see
text and Tab.~\ref{tab:SD:Hg}). The same color code as in
Fig.~\ref{fig:Pufission} is used.
}
\end{figure}

For the even-even nuclei whose energy curves are shown in
Fig.~\ref{fig:Hg:all:def:SLy5s1},
the actual values of the excitation energies of the SD minimum and
corresponding proton deformation parameters are summarized in
Fig.~\ref{fig:Hgsuper} for all eight parametrizations. A smaller value
of $a_{\rm surf}$ systematically yields a smaller excitation energy.
The deformation parameters are very similar with only small
variations. There is the trend that parametrizations with
lower $a_{\rm surf}$ produce slightly larger $\beta_{20}$ than the
parametrizations with higher $a_{\rm surf}$. For $\beta_{40}$ and
$\beta_{60}$, however, there is no such general trend.

\begin{table}[t!]
\caption{\label{tab:sd}\label{tab:SD:Hg}
Charge quadrupole deformation $\beta_{2,p}$ and excitation energies
$\Delta E_{\text{SD}}$
(in MeV) of the superdeformed minima of Hg isotopes as indicated.
Experimental data on the $\beta_{2,p}$ are taken from~\cite{Singh02},
whereas the estimates for $\Delta E_{\text{SD}}$ are taken from
\cite{Wilson10} ($^{190}$Hg),
\cite{Lauritsen00} ($^{192}$Hg), and
\cite{Khoo96} ($^{194}$Hg; the authors do not provide an
estimate of the error bar).
}
\begin{ruledtabular}
\begin{tabular}{llllllll}
\noalign{\smallskip}
 & \multicolumn{2}{c}{$^{190}$Hg} &
   \multicolumn{2}{c}{$^{192}$Hg} &\multicolumn{2}{c}{$^{194}$Hg} \\
\noalign{\smallskip}
\cline{2-3}\cline{4-5}\cline{6-7}
\noalign{\smallskip}
 & $\beta_{2,p}$ & $\Delta E_{\text{SD}}$
 & $\beta_{2,p}$ & $\Delta E_{\text{SD}}$
 & $\beta_{2,p}$ & $\Delta E_{\text{SD}}$  \\
\noalign{\smallskip}
\hline
\hline
\noalign{\smallskip}
SLy5s8 & 0.65 & 4.12 & 0.65 & 5.12 & 0.64 & 6.39 \\
SLy5s7 & 0.65 & 3.99 & 0.65 & 4.98 & 0.64 & 6.27 \\
SLy5s6 & 0.65 & 3.86 & 0.65 & 4.83 & 0.64 & 6.14 \\
SLy5s5 & 0.65 & 3.73 & 0.65 & 4.69 & 0.64 & 6.01 \\
SLy5s4 & 0.66 & 3.61 & 0.65 & 4.55 & 0.64 & 5.87 \\
SLy5s3 & 0.66 & 3.50 & 0.66 & 4.41 & 0.64 & 5.73 \\
SLy5s2 & 0.66 & 3.38 & 0.66 & 4.27 & 0.65 & 5.58 \\
SLy5s1 & 0.66 & 3.26 & 0.66 & 4.12 & 0.65 & 5.42 \\
\noalign{\smallskip}\hline\noalign{\smallskip}
Exp.  & 0.61(9) & 4.53(2) & 0.70(4) & 5.3(5) & 0.58(3) & 6.017 \\
\noalign{\smallskip}
\end{tabular}
\end{ruledtabular}
\end{table}

Figure~\ref{fig:Hgsuper} also shows experimental data where available.
Their detailed comparison with calculated values is provided by
Tab.~\ref{tab:sd}. As already mentioned, the SD bands decay out
to ND states at finite spin, such that the band head remains unobserved.
The charge quadrupole moments are taken from the compilation of
Ref.~\cite{Singh02} and were each obtained from transition quadrupole
moments averaged over several transitions between excited states in the
yrast SD band. The excitation energies are estimated from
$\gamma$-ray energies of transitions linking SD and ND states and
subsequent extrapolation of the SD band to zero spin. For $^{190}$Hg and
$^{194}$Hg, the latter has been achieved following discrete transitions
\cite{Khoo96,Wilson10}, whereas the estimate for $^{192}$Hg stems from
a statistical analysis of a quasi-continuous spectrum \cite{Lauritsen00}.

For all three nuclei and all parametrizations, the calculated $\beta_{2,p}$
are almost identical and agree well, within error bars, with the empirical
values with the exception of $^{194}$Hg for which it is slightly
overestimated. The excitation energy $\Delta E_{\text{SD}}$, however, is
less well described. First, the calculated values increase too quickly with
$N$, and second, their values are better described for the parametrizations
with large $a_{\text{surf}}$, which has to be contrasted with the discussion
of the fission barriers of $^{180}$Hg, $^{226}$Ra, and $^{240}$Pu above,
for which the barriers were best described by the parametrizations with
lowest $a_{\text{surf}}$. This, however, might be a trivial consequence
of the slope of the $\Delta E_{\text{SD}}$ being incorrect as a function
of $N$.

\begin{figure}
\includegraphics[width=8.0cm]{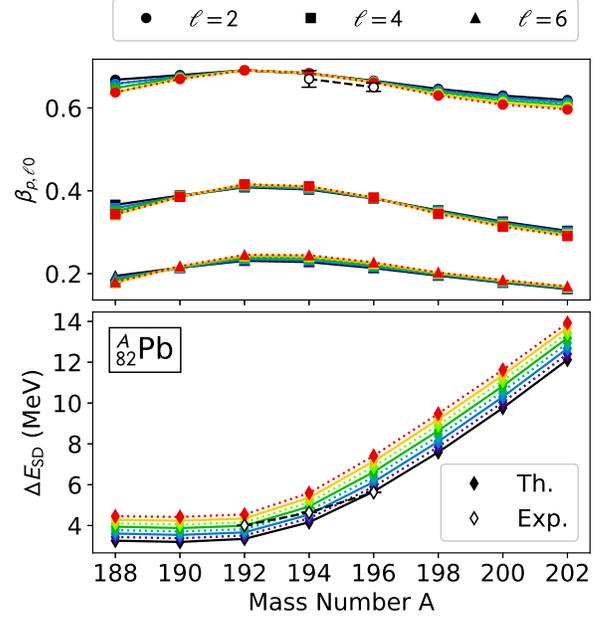}
\caption{\label{fig:Pbsuper}
Same as Fig.~\ref{fig:Hgsuper}, but for Pb isotopes (see also
Tab.~\ref{tab:SD:Pb}).
}
\end{figure}

The same overall pattern is also found when looking to such data for
the adjacent even-even $^{192-196}$Pb isotopes, see Fig.~\ref{fig:Pbsuper}
and Table~\ref{tab:SD:Pb}.
The calculated quadrupole moments are near-identical for all parametrizations
and nuclei and agree well with data, whereas the excitation energy
$\Delta E_{\text{SD}}$ increases much too quickly with $N$. As a consequence,
the data for $^{192}$Pb are best described with a parametrization with large
$a_{\text{surf}}$, whereas for $^{196}$Pb a low $a_{\text{surf}}$ is needed,
with $^{194}$Pb falling in between.

A possible incorrect trend in the surface symmetry energy coefficient
$a_{\text{ssym}}$ is not likely to be the explanation of the wrong trend
of excitation energies. First of all, the deviation from experiment changes
too quickly with $N$. Second, as argued when comparing barrier heights
of $^{180}$Hg, $^{226}$Ra and $^{240}$Pu, the isospin dependence of the
surface energy has probably to be changed in the opposite direction, which
would make the disagreement with data for the SD minima in the $A \approx 190$
region even worse.

\begin{table}[b!]
\caption{\label{tab:SD:Pb}
Same as Table~\ref{tab:SD:Hg}, but for the Pb isotopes as indicated.
Experimental data for $\beta_{2,p}$ are taken again from~\cite{Singh02},
whereas estimates for $\Delta E_{\text{SD}}$ are taken from
\cite{Wilson03} ($^{192}$Pb),
\cite{Hauschild97} ($^{194}$Pb), and
\cite{Wilson05} ($^{196}$Pb).
}
\begin{ruledtabular}
%\begin{tabular}{l@{\hspace{2em}}ll@{\hspace{2em}}ll@{\hspace{2em}}ll}
\begin{tabular}{lllllll}
\noalign{\smallskip}
 & \multicolumn{2}{c}{$^{192}$Pb}
 & \multicolumn{2}{c}{$^{194}$Pb}
 & \multicolumn{2}{c}{$^{196}$Pb} \\
\noalign{\smallskip}
\cline{2-3}\cline{4-5}\cline{6-7}
\noalign{\smallskip}
 & $\beta_{2,p}$ & $\Delta E_{\text{SD}}$
 & $\beta_{2,p}$ & $\Delta E_{\text{SD}}$
 & $\beta_{2,p}$ & $\Delta E_{\text{SD}}$  \\
\noalign{\smallskip}
\hline
\hline
\noalign{\smallskip}
SLy5s8 & 0.69 & 4.52 & 0.68 & 5.58 & 0.66 & 7.41 \\
SLy5s7 & 0.69 & 4.30 & 0.68 & 5.35 & 0.66 & 7.14 \\
SLy5s6 & 0.69 & 3.11 & 0.68 & 5.13 & 0.66 & 6.88 \\
SLy5s5 & 0.69 & 2.96 & 0.68 & 4.92 & 0.66 & 6.62 \\
SLy5s4 & 0.69 & 2.80 & 0.68 & 4.72 & 0.66 & 6.37 \\
SLy5s3 & 0.69 & 2.65 & 0.68 & 4.53 & 0.66 & 6.13 \\
SLy5s2 & 0.69 & 2.49 & 0.68 & 4.34 & 0.66 & 5.90 \\
SLy5s1 & 0.69 & 2.34 & 0.68 & 4.15 & 0.67 & 5.66 \\
\noalign{\smallskip}\hline\noalign{\smallskip}
Exp.  &  & 4.01 & 0.67(2) & 4.64 & 0.65(1) & 5.63(1) \\
\noalign{\smallskip}
\end{tabular}
\end{ruledtabular}
\end{table}

A more likely explanation for the incorrect description of the slope
of the $\Delta E_{\text{SD}}$ are shell effects, that probably are not only
incorrect in total size but also isotopic dependence. As argued above, the
net shell effects are not the same for all parametrizations. The question of
which one describes the shell structure best is difficult to answer from the
material analyzed here and therefore beyond the scope of out study. There are,
however, many indications that a satisfactory simultaneous description of
nuclear bulk properties and shell structure cannot be achieved with the
presently used functional form of the nuclear EDF
\cite{unedf2,Lesinski07a,Bonneau07,Bender09a,Schunck10a}.

A discussion of the excitation energies of the SD band heads in the
$A \approx 190$ region calculated with a large number of traditional
parametrizations of the Skyrme EDF can be found in Ref.~\cite{Nikolov11a}.
Results shown there indicate that the too quick increase of calculated
$\Delta E_{\text{SD}}$ values is a virtually universal feature of
existing Skyrme interactions, although with differences in magnitude.

\begin{figure}[t!]
\includegraphics[width=8.0cm]{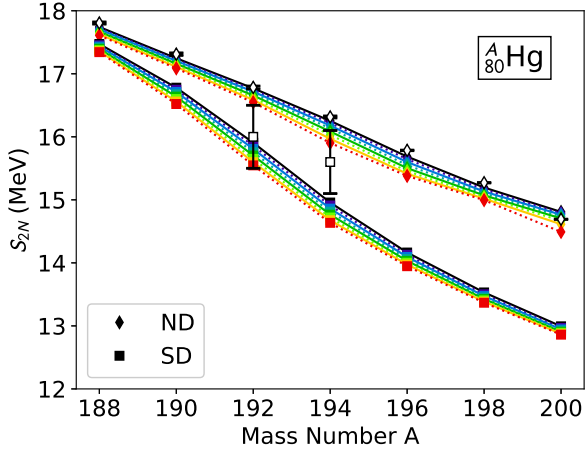}
\caption{\label{fig:s2n:hg}
Two-neutron separation energies $S_{2n}$ of ground states and superdeformed
states of neutron-deficient Hg isotopes as indicated using the same color
code as in Fig.~\ref{fig:Pufission} for the calculated values. Experimental
data are represented by open white symbols. For ground states they are taken
from \cite{AME16}, for SD states from \cite{Wilson10}.
}
\end{figure}

\begin{figure}[t!]
\includegraphics[width=8.0cm]{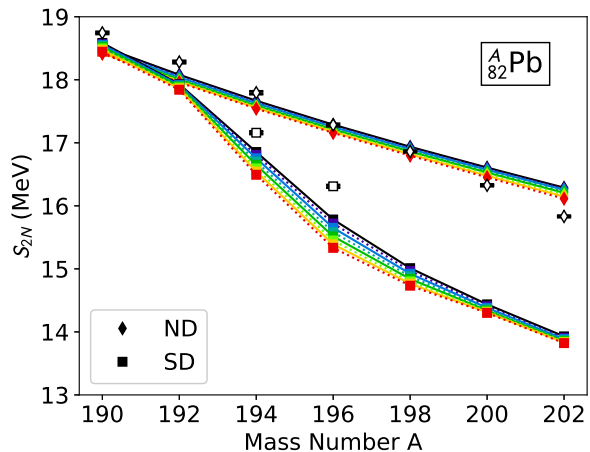}
\caption{\label{fig:s2n:pb}
Same as Fig.~\ref{fig:s2n:hg}, but for Pb isotopes.
}
\end{figure}

The slope of the excitation energies can also be analyzed through the
two-neutron separation energy calculated between SD minima
\begin{eqnarray}
\label{eq:S2n:SD}
\lefteqn{S_{2n,\text{SD}}(N,Z)
} \nn \\
& = & E_{\text{SD}}(N-2,Z) - E_{\text{SD}}(N,Z)
      \nn \\
& = & \big[ E(N-2,Z) + \Delta E_{\text{SD}}(N-2,Z) \big]
      \nn \\
&   &
     -\big[ E(N,Z) + \Delta E_{\text{SD}}(N,Z) \big]
      \nn \\
& = & S_{2n}(N,Z) + \Delta E_{\text{SD}}(N-2,Z)
                 - \Delta E_{\text{SD}}(N,Z) \, , \nonumber \\
\end{eqnarray}
where $\Delta E_{\text{SD}}(N,Z)$ is the (positive) excitation energy
of the SD minimum and $E(N,Z)$ the (negative) binding energy of the
nuclide $(N,Z)$. This quantity has been discussed earlier in
Refs.~\cite{Heenen98,Wilson10}.
Results for the $S_{2n,SD}(N,Z)$ of neutron-deficient Hg and Pb isotopes
are compared with the conventional two-neutron separation energies between
the ground states $S_{2n}(N,Z)$ and experimental data for both where
available in Figs.~\ref{fig:s2n:hg} and~\ref{fig:s2n:pb}.

The authors of Ref.~\cite{Wilson10} have argued that, compared to the
two-particle separation energies of the respective ground states,
``Naively, one might expect a reduction in both $S_{2n}$ and $S_{2p}$ at
superdeformation, since both the binding energies per nucleon and the
Coulomb barrier are lower.''
This, however, is a fallacy. First of all, if the SD states were all at
constant excitation energy, then it is obvious that their two-particle
separation energies would be identical to those of the ground states
irrespective of what the $E/A$ and Coulomb barrier are, as the latter
are reduced for both nuclei entering the calculation of the separation
energy. Second, if the $S_{2n}$ of SD states were systematically smaller
than those of ND states in a given isotopic chain,
$S_{2n,SD}(N,Z) < S_{2n}(N,Z)$, then Eq.~\eqref{eq:S2n:SD} implies that the
excitation energy of the SD states would always systematically increase with $N$, i.e.\
$\Delta E_{\text{SD}}(N-2,Z) < \Delta E_{\text{SD}}(N,Z)$. While the latter
is indeed the case for the few $A \approx 190$ isotopes for which there
are data, the systematics of data across the entire nuclear chart does
not follow such rule.

Irregularities in the systematics of two-nucleon separation energies can
either indicate a gap in the single-particle spectrum or a large change in
correlations \cite{Bender08}, such as for example a large change in
deformation.

As can be seen from Figs.~\ref{fig:s2n:hg} and~\ref{fig:s2n:pb},
all SLy5sX give a fair description of the $S_{2n}$ of the ground
states in both isotopic chains. For Pb isotopes, the calculated $S_{2n}$
do not fall off with exactly the same slope as the data, though.
Differences between the parametrizations are very small in both slope
and offset, which in view of their systematically different nuclear
matter properties, see Table~\ref{tab:INM}, could not necessarily be
expected. This point is analyzed further in
Sect.~\ref{subsect:isotopic} below. Similarly, differences between the
$S_{2n,\text{SD}}(N,Z)$ calculated with different parametrizations
remain small, with an offset between the curves that is nearly
identical to the one between the $S_{2n}$. In the region around
$A \simeq 194$, the $S_{2n,\text{SD}}$
are smaller than the $S_{2n}$ and also falling off much quicker. For
smaller mass numbers, the curves approach each other and become parallel.
This behavior is simply a consequence of the up-bend of the
$\Delta E_{\text{SD}}$ with increasing mass in both isotopic chains
as illustrated by Figs.~\ref{fig:Hgsuper} and~\ref{fig:Pbsuper}.

The drop of the $S_{2n,\text{SD}}$ of Pb isotopes when going from $^{194}$Pb
to $^{196}$Pb is also reasonably described, although the overall size of the
$S_{2n,\text{SD}}$ is underestimated. Comparison with data for the Hg isotopes
is compromised by their large error bars, but their values are
also tentatively underestimated. In both cases, this is a direct
consequence of the too large up-bend of the calculated $\Delta E_{\text{SD}}$.

%=========================================================================

\subsection{Superdeformed rotational bands}
\label{sect:SD:rot}

\begin{figure}[t!]
\includegraphics[width=8.8cm]{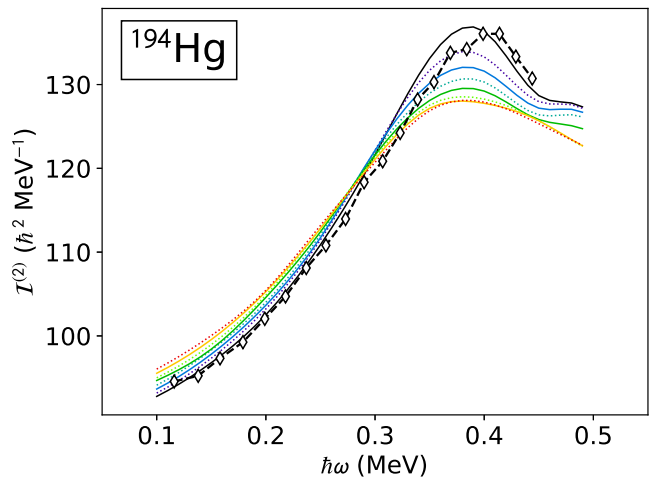}
\caption{\label{fig:rotband}
Dynamical moment of inertia $\mathcal{I}^{(2)}$ of the superdeformed
band in $^{194}$Hg as a function of cranking frequency $\hbar \omega$ for the
SLy5sX parametrizations indicated by the same color code as in
Fig.~\ref{fig:Pufission}. The white diamonds represent experimental data
for the yrast SD-1 band taken from Ref.~\cite{Singh02}.
}
\end{figure}

The properties of the $0^+$ band-heads of the SD yrast bands are obtained
from extrapolation. More direct experimental information on SD states in
the $A \approx 190$ region is provided by the rotational bands.
Figure~\eqref{fig:rotband} displays the dynamical moment of inertia
\begin{equation}
\mathcal{I}^{(2)}
\equiv \left( \frac{d^2 E}{dJ^2} \right)^{-1}  \, ,
\end{equation}
of the yrast SD-1 band of $^{194}$Hg as a function of rotational frequency
$\omega$
\begin{equation}
\omega
\equiv \frac{dE}{dJ}
\end{equation}
as obtained from cranked HFB calculations assuming triaxial shapes as
described in Refs.~\cite{Gall1994,Rigollet1999,Hellemans12}. In such
calculations one minimizes the Routhian
\begin{equation}
\label{eq:routhian}
R = E - \omega_{\rm th} \langle \hat{J}_{z} \rangle \, .
\end{equation}
The constraint on the expectation value of the $z$ component of angular
momentum breaks intrinsic time-reversal invariance, such that the time-odd
terms of Eq.~\eqref{eq:SkTodd} contribute to the EDF and the single-particle
Hamiltonian. The calculations described here have been performed using a
linear constraint with fixed $\omega_{\rm th}$, the value of which is also
used to draw the curves in Fig.~\ref{fig:rotband}.

For an even-even nucleus and $\Delta J = 2 \, \hbar$ transitions, the
experimental values for $\omega$ and $\mathcal{I}^{(2)}$ are calculated
from data provided by Ref.~\cite{Singh02} with the help of finite
difference formulas \cite{Dudek92}
\begin{eqnarray}
\hbar\omega_{\rm exp}
& = & \tfrac{1}{4}
      \big[ E_{\gamma} (J + 2 \rightarrow J) + E_{\gamma} (J \rightarrow J-2)
      \big] \, ,
      \\
\mathcal{I}^{(2)}_{\rm exp}
& = & \frac{4}
           {E_{\gamma} (J + 2 \rightarrow J) - E_{\gamma} (J \rightarrow J-2)}
      \, ,
\end{eqnarray}
where $E_{\gamma}$ is the energy of the $\gamma$ ray emitted in the
transition from the level with $J$ to the level with $J-2$. Both can
be determined without knowing the angular momentum $J$ of the states
involved as is often the case for SD rotational bands.

All SLy5sX parametrizations give a fair description of the data; in
particular those with the smallest surface coefficients perform as well as
the best parametrizations identified in previous
applications~\cite{Terasaki95,Hellemans12}. The differences
between the parametrizations observed here are in general smaller in size
than the ones obtained when varying the strength of the pairing
interaction or the not well-constrained time-odd terms of the
EDF~\cite{Hellemans12}, two ingredients of the EDF that are much
smaller in magnitude than the surface energy.

There are small systematic differences between the parametrizations; at low
frequency, SLy5s8 exhibits the largest moment of inertia, while at higher
frequencies the ordering is reversed and SLy5s1 has the largest value of
$\mathcal{I}^{(2)}$. The difference between the parametrizations is largest
around $\hbar \omega \approx 0.38$ MeV, where the moment of inertia first
plateaus and later drops off. In this region, SLy5s1 follows the data
most closely. This, however, is not related to any difference in deformation
between the parametrizations, which become even closer with increasing $J$.
In fact, this saturation of the rotational band in this region
is usually associated with the alignment of neutron intruder
orbitals~\cite{Terasaki95}, which are at a slightly different position for
each Sly5sX because of the systematic differences in spin-orbit strength,
see the discussion of Fig.~\ref{fig:Hg:Nilsson}.

%
%--------------------------------------------------------------------------
%
\subsection{Isotopic trends of ground states}
\label{subsect:isotopic}

\begin{figure}[t!]
\includegraphics[width=8cm]{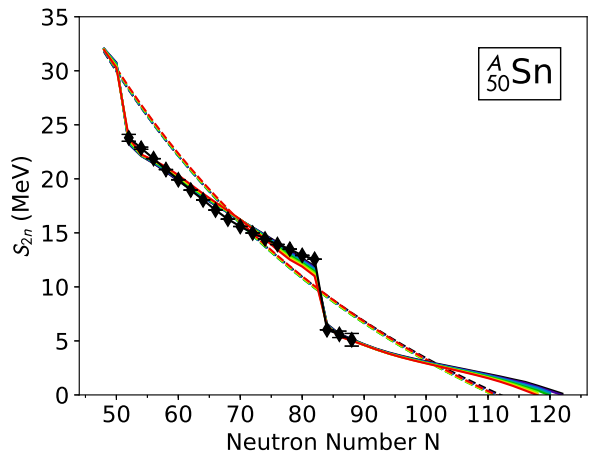}
\caption{\label{fig:sn:s2n}
Two-neutron separation energies $S_{2n}$ of even-even Sn isotopes. Solid
lines are results from HFB calculations with the parametrizations as
indicated, whereas dotted lines are the $S_{2n,\text{LDM}}$ calculated
through Eq.~\eqref{eq:S2n:LDM} from LDM models with same INM and SINM
parameters as the SLy5sX. Experimental data are taken from \cite{AME16}.
}
\end{figure}

In the discussion of Figs.~\ref{fig:s2n:hg} and~\ref{fig:s2n:pb}
we have seen that the SLy5sX parametrizations give similar results
for two-neutron separation energies. In view of the systematic differences
between INM properties discussed in Section~\ref{sec:differences} this
might come as a surprise. Indeed, as listed in Table~\ref{tab:INM}, the
symmetry energy coefficient $a_{\text{sym}}$ steadily changes from
31.43~MeV for SLy5s1 to 32.64~MeV for SLy5s8, which has no
significant impact on the $S_{2n}$ of neutron-deficient Hg and Pb isotopes.

The same result is found when looking at the $S_{2n}$ along the chain of Sn
isotopes, Fig.~\ref{fig:sn:s2n}. The curves for the various parametrizations
almost everywhere fall on top of each other, with the exception of nuclei just
below the $N=82$ shell closure and the least bound ones close to
the neutron drip line. The global trend of the data is well described,
including the size of the jump at $N=82$.

For comparison, in Fig.~\ref{fig:sn:s2n} we also show the two-neutron
separation energy obtained as
\begin{equation}
\label{eq:S2n:LDM}
S_{2n,\text{LDM}}(N,Z)
\equiv - 2 \frac{d E_{\text{LDM}}(N,Z)}{d N}
\end{equation}
from the LDM energy~\eqref{eq:mac} calculated from the INM and SINM properties
of the SLy5sX parametrizations. Again, the curves almost perfectly fall
on top of each other. Qualitatively, the same is found when looking at the
two-proton separation energies of $N=82$ isotones, cf.~Fig.~\ref{fig:n82:s2p}.
Several comments on these results are in order
\begin{enumerate}[(i)]
\item
While the evolution of observables along isotopic and isotonic
chains is frequently used for the analysis of the impact of isovector terms,
which scale with the square of the asymmetry $I=N-Z$, the
mass number $A=N+Z$ also changes at the same time, such that isoscalar and
isovector effects are inevitably intertwined.

\begin{figure}[t!]
\includegraphics[width=7.0cm]{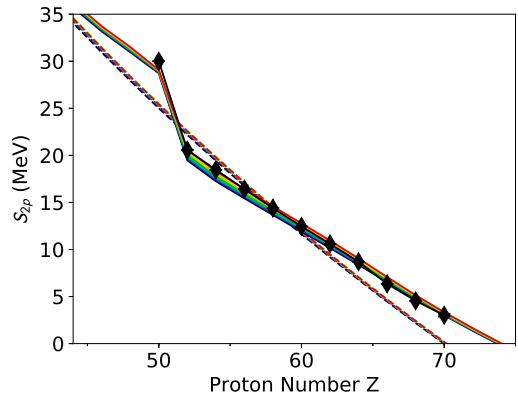}
\caption{\label{fig:n82:s2p}
Two-proton separation energies $S_{2p}$ of even-even $N=82$ isotones
represented in the same way as the  $S_{2n}$ of Sn isotopes in
Fig.~\ref{fig:sn:s2n}.
}
\end{figure}

\item
For the SLy5sX parametrizations, both the isoscalar and isovector INM
parameters are changing as a function of the constraint on $a_{\text{surf}}$,
cf.\ Table~\ref{tab:INM}. The coefficients $a_{\text{vol}}$ and
$a_{\text{sym}}$ of the dominating contributions to the LDM energy change
in opposite direction. Combined with the changes in the surface terms, the net
effect is that the separation energies are almost unchanged when going
from one SLy5sX parametrization to another. This would be different if
all other LDM parameters were fixed and only $a_{\text{sym}}$ varied.

\item
While the HFB results for the $S_{2n}$ and $S_{2p}$ oscillate around the LDM
results, at no point do the HFB values (or the experimental data, for that
matter) follow the slope of the LDM values.

\end{enumerate}
Indeed, the two-neutron separation energy $S_{2n} \approx - 2 \lambda_n$
approximates the negative of twice the Fermi energy of neutrons, and similar
for protons. As a consequence, the overall trend of the microscopically
calculated $S_{2n}$ and $S_{2p}$ is determined by the position of
single-particle levels and the correlations from pairing and deformation
modes. These are very similar for all SLy5sX parametrizations except
for the isotopes just below $^{132}$Sn, where the position of the
$h_{11/2^-}$ intruder neutron level depends strongly on the parametrization,
very similar to what has been found for the proton $h_{11/2^-}$
level of $^{188}$Hg in Fig.~\ref{fig:Hg:Nilsson}.

The impact of the variations in  $a_{\text{vol}}$ and $a_{\text{sym}}$ can
only be seen when looking at observables as a function of $A$ at constant $I$
and as a function of $I$ at constant $A$, respectively. For example, the
softening of $E/A$ with increasing mass number when going from SLy5s1 to
SLy5s8 visible in Fig.~\ref{fig:SHE:LDM} is a consequence of the increase
of $a_{\text{vol}}$ from $-15.77$~MeV to $-16.10$~MeV.

This has some visible influence on the global trend of mass residuals of
heavy nuclei. Changing $a_{\text{vol}}$ by as little as 0.1~MeV
while keeping all other LDM coefficients constant changes the binding
energy of $^{208}$Pb by 20.8~MeV and that of $^{240}$Pu by 24~MeV. Like
many other widely-used parametrizations of the Syrme EDF, all SLy5sX
underbind actinide nuclei, but to a different degree. Going from SLy5s8
to SLy5s1, the underbinding of $^{240}$Pu increases from about 6~MeV to
almost 16~MeV, which is close to what can be naively expected from the change
in $a_{\text{vol}}$, provided that $^{208}$Pb has the same binding energy
in both cases. Because of the change of all terms in Skyrme EDF
\eqref{eq:skyrme:energy} and the corresponding LDM energy \eqref{eq:mac},
however, this estimate is less reliable than it may seem.

%
%--------------------------------------------------------------------------
%
\subsection{Shape coexistence at normal deformation}
\label{sec:normaldef}

%- - - - - - - - - - - - - - - - - - - - - - - - - - - - - - - - - - - - -
%
\subsubsection{Shape coexistence in $^{186}$Pb}

Like the neutron-deficient Hg isotopes, Pb isotopes
in the $A \approx 180$ region exhibit shape coexistence at normal deformation.
A spectacular example is given by $^{186}$Pb, for which the three lowest-lying
states are $0^+$ states interpreted as spherical, prolate and oblate
shapes \cite{Andreyev00}. The ground state is associated with a spherical
shape, and the $0^+$ states at 536(21) and 655(21)~keV are oblate and prolate
configurations. Correlations do of course lead to a mixture of the pure
configurations.

\begin{figure}[t!]
\includegraphics[width=7.8cm]{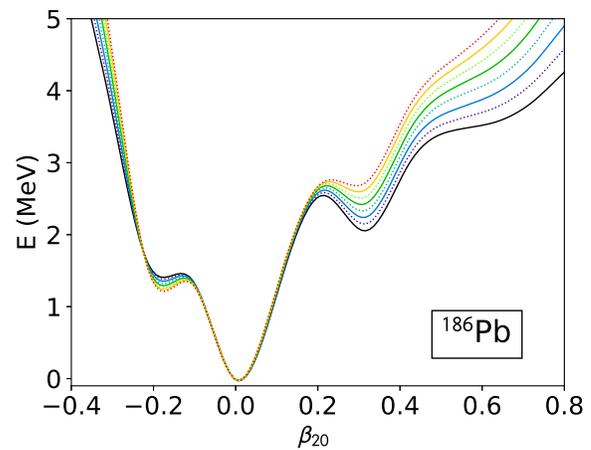}
\caption{\label{fig:pb186}
Deformation energy of $^{186}$Pb as a function of the axial mass
quadrupole deformation $\beta_{20}$ for the SLy5sX parametrizations
indicated with the same color code as in Fig.~\ref{fig:Pufission}.
}
\end{figure}

This behavior is qualitatively reproduced by all SLy5sX parametrizations,
see Fig.~\ref{fig:pb186}. The excitation energy of the deformed minima,
however, is largely overestimated.  The situation can be improved by
reducing the pairing strength \cite{Bender04b}, although this would in turn
degrade the description of other observables, most notably the dynamical
moment of inertia of the SD band of $^{194}$Hg of Fig.~\ref{fig:rotband}.

In any event, on the scale of the Figure, changing $a_{\text{surf}}$ has
no visible impact on the deformation around the spherical point up to
oblate deformations well beyond the minimum, which is a consequence of
a simultaneous change in shell effects as deduced already from
Figs.~\ref{fig:Hg180:diff}, \ref{fig:HgLDM}, and~\ref{fig:Hg194:diff}
for adjacent nuclei. By contrast, reducing $a_{\text{surf}}$
slightly lowers the prolate minimum and makes it more pronounced. The impact
of changing $a_{\text{surf}}$, however, only becomes clearly visible at
larger deformation than the one of the coexisting states. Altogether,
this indicates that the surface energy coefficient is not the most
relevant degree of freedom for the fine-tuning of shape-coexisting states
in neutron-deficient Pb isotopes. Modifying the pairing strength or
changing the shell structure by variation of the tensor terms has a much
larger effect \cite{Bender09a} on the energy differences between the
minima.\footnote{According to the tables provided by~\cite{Jodon16}, with
$a_{\text{surf}}^{\text{MTF}} \simeq 19.0 \pm 0.25$ MeV, the surface energy
coefficients of the Tij parametrizations used in Ref.~\cite{Bender09a}
are all on the upper end of the scale covered by the SLy5sX.}

It has to be noted that these states are much better described in
beyond-mean-calculations that combine projection on angular momentum and
particle-number with configuration mixing in a generator coordinate method
(GCM) using parametrizations that, at the mean-field level, perform similar
to the SLy5sX with low $a_{\text{surf}}$ \cite{Rodriguez04a,Bender04b,Duguet03}.
This indicates that a mean-field description of shape coexistence in this
mass region might not be sufficient, as each of the three low-lying $0^+$
states gains a quite different amount of correlation energy.

%- - - - - - - - - - - - - - - - - - - - - - - - - - - - - - - - - - - - -
%
\subsubsection{Shape coexistence in $^{74}$Kr}
\begin{figure}
\includegraphics[width=8.8cm]{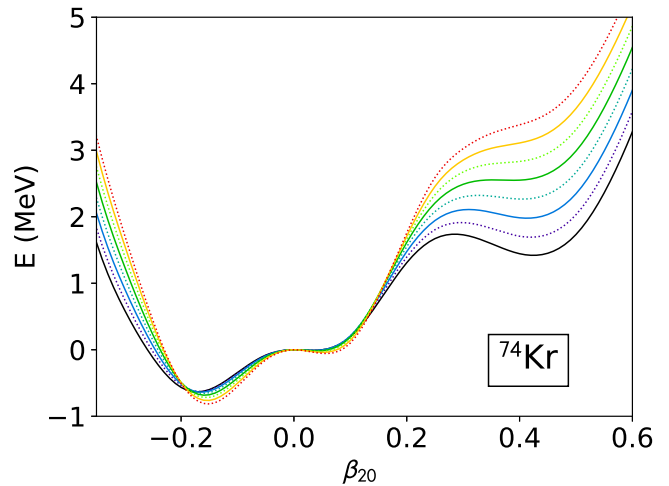}
\caption{\label{fig:kr74}
Same as Fig.~\ref{fig:pb186}, but for $^{74}$Kr.
}
\end{figure}

Another region of the nuclear chart for which detailed data on shape
coexistence are available is the neutron-deficient Kr region. As an
example, Fig.~\ref{fig:kr74} shows the deformation energy curve for axial
states of $^{74}$Kr. At large deformation $| \beta_{20} | \gtrsim 0.25$,
the curves behave as expected from the $a_{\text{surf}}$ of each
parametrization, but at small deformations their order is inverted: SLy5s8
gives a softer energy curve around the spherical point than SLy5s1. The
reason is again the difference in spin-orbit interaction, which is weaker
for the former than for the latter. The resulting small changes in shell
structure then increase or decrease the binding energy for near-spherical
shapes relative to those at larger deformation. What appears as a prolate
shoulder in calculations with SLy5s8 becomes a pronounced low-lying excited
minimum with SLy5s1.

Electromagnetic transition moments between the low-lying states, however,
indicate that the ground-state band is dominated by prolate shapes with
$\beta_{2} \approx 0.4$
\cite{Clement07}, whereas an excited band is mainly built from oblate
shapes. From this follows that none of the SLy5sX describes the correct
relative position of the minima. Extrapolating the trend of the SLy5sX,
$a_{\text{surf}}$ had to be reduced by more than 1~MeV below the value
of SLy5s1 in order to get the correct order of the minima. Such
parametrization would have unrealistic properties for fission barriers
of actinides. However, for this nucleus that is almost on the $N=Z$ line,
the combined reduction of $a_{\text{surf}}$ by about 0.6~MeV and (for this
nucleus irrelevant) increase of $a_{\text{ssym}}$ proposed in
Section~\ref{sect:correlation} in order to obtain simultaneously the
fission barrier heights of $^{180}$Hg and $^{240}$Pu would nevertheless
improve on the situation.

A low value of $a_{\text{surf}}$ is most probably a necessary ingredient
for the quantitative description of shape coexistence in this region of
the chart of nuclei, but it can again be argued that other aspects of
the parametrizations such as shell structure and pairing correlations
are of at least equal importance and require further fine-tuning in order
to describe the energy difference between the minima in a mean-field
calculation.

Like for $^{186}$Pb, beyond-mean-field calculations based on
parametrizations that give a similar energy surface as SLy5s1 give
an excellent description of low-lying states in this
nucleus \cite{Girod09a,Rodriguez14a,Yao14a}, hinting again at the possible
insufficiency of mean-field calculations to quantitatively describe shape
coexistence phenomena at normal deformation.

%- - - - - - - - - - - - - - - - - - - - - - - - - - - - - - - - - - - - -
%
\subsubsection{Structure of $^{110}$Zr}

Nuclei in the Zr region exhibit a rich and quickly evolving structure,
that is notoriously difficult to describe in all details by mean-field
models \cite{Bender09a}. Data suggest that $^{80}$Zr is prolate deformed,
while $^{90}$Zr is quasi doubly-magic. The heavier $^{96}$Zr is usually
interpreted as a spherical nucleus with low-lying collectively deformed states,
while adding four neutrons leads to the well prolate deformed $^{100}$Zr.
A model-dependent analysis of recent data suggest that the even heavier
neutron-rich $^{110}$Zr is $\gamma$-soft with a preference for prolate
shapes~\cite{Paul17}.

\begin{figure}[t!]
\includegraphics[width=8.0cm]{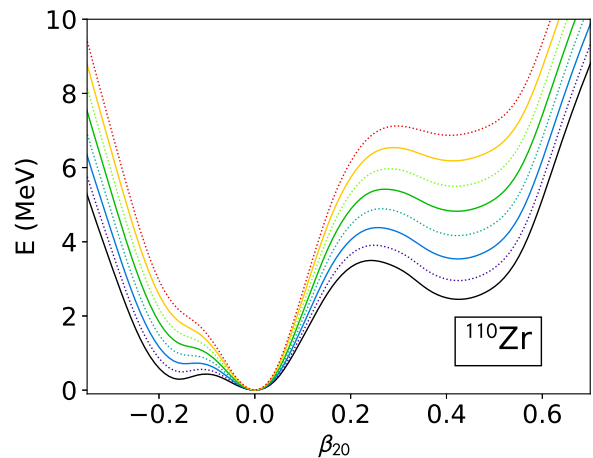}
\caption{\label{fig:zr110}
Same as Fig.~\ref{fig:pb186}, but for $^{110}$Zr.
}
\end{figure}

The axial energy curves as obtained with the SLy5sX parametrizations for this
last nucleus are shown in Fig.~\ref{fig:zr110}. Going from SLy5s8 to SLy5s1,
the energy surface becomes softer, with an oblate minimum developing at
$\beta_{20} \simeq -0.1$ that is lowered to 110~keV excitation energy
for SLy5s1,  while prolate states at $\beta_{20} \simeq 0.4$ are
lowered by about 5~MeV. The absolute minimum, however, is spherical for all
parametrizations, with the prolate states remaining at excitation energies
of 2.34~MeV for SLy5s1 up to 6.48~MeV for SLy5s8.

\begin{figure}
\centerline{\includegraphics[width=5.8cm]{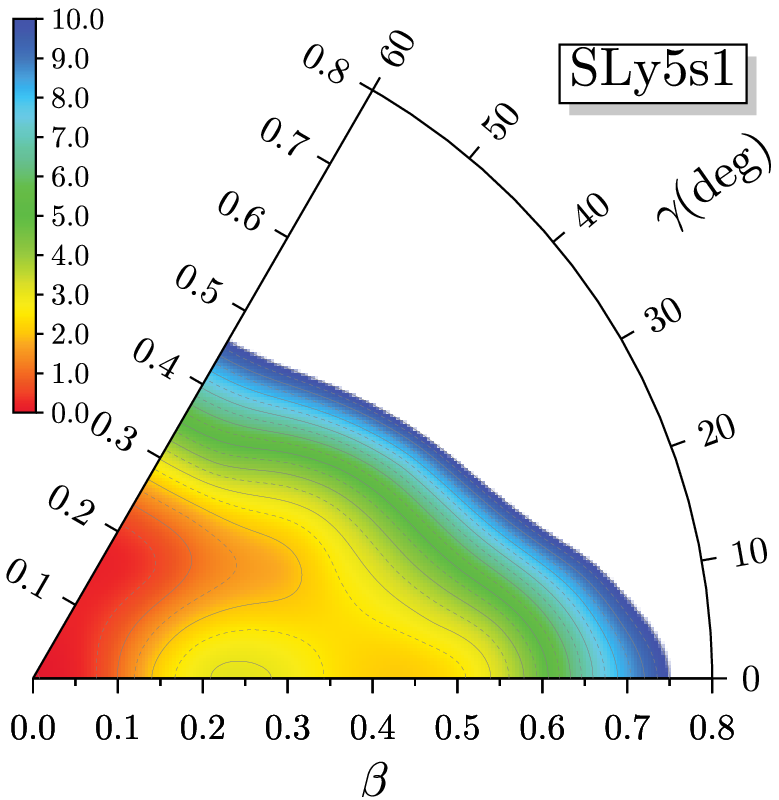}}
\centerline{\includegraphics[width=5.8cm]{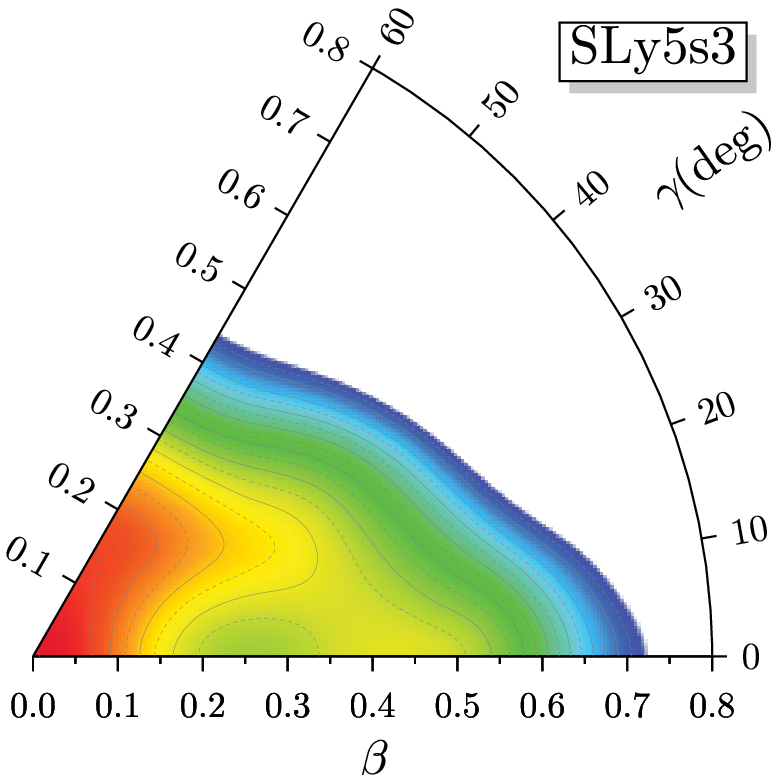}}
\centerline{\includegraphics[width=5.8cm]{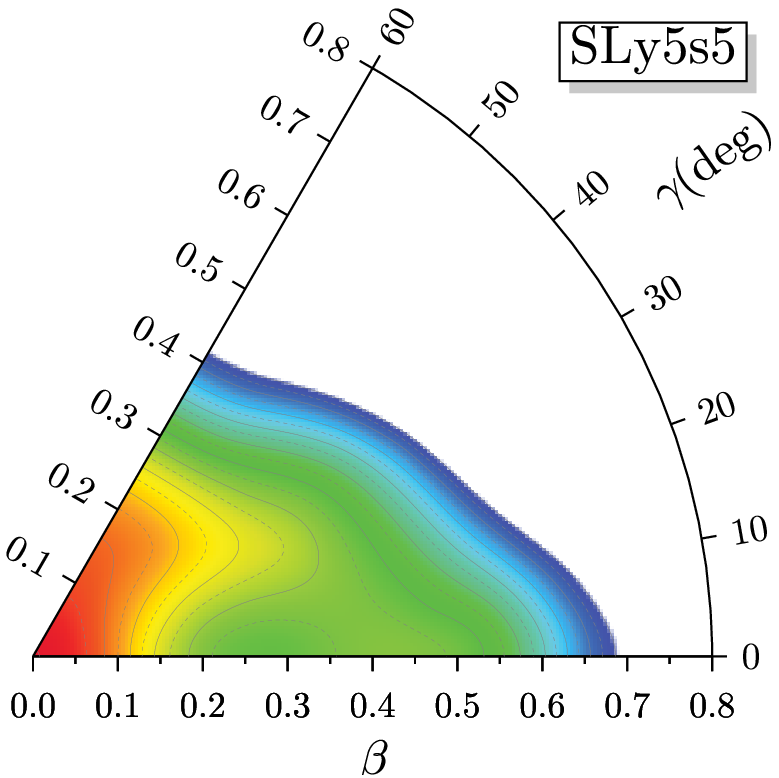}}
\centerline{\includegraphics[width=5.8cm]{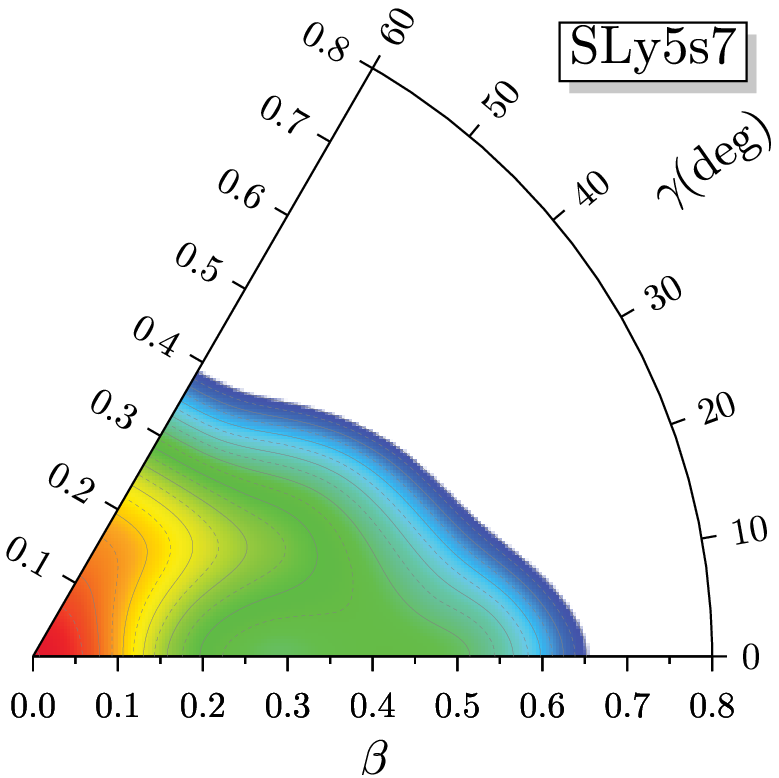}}
\caption{\label{fig:Zr110:beta:gamma}
Deformation energy surfaces in the $\beta$-$\gamma$ plane
of $^{110}$Zr obtained with the SLy5sX parametrizations as indicated.
}
\end{figure}

As can be seen from the energy surfaces in the full $\beta$-$\gamma$ plane,
Fig.~\ref{fig:Zr110:beta:gamma}, the prolate structure is in fact a saddle
for all SLy5sX parametrizations. The D1S parametrization of the Gogny force
gives an energy surface that is very similar to the one of SLy5s1.
As has been demonstrated in Ref.~\cite{Paul17}, a modified D1S with increased
strength of the spin-orbit interaction (from a value slightly larger than the
one for SLy5s1 to an even larger one) significantly
improves the description of the available spectroscopic data. The main effects
of this change are that the spherical state is slightly pushed up and that the
global minimum becomes prolate. A similar effect can be obtained by a
modification of tensor terms \cite{Bender09a}, although this might degrade
properties of other nuclei.

Without discussing them in detail, we can mention that the changes of the
deformation energy curves of $^{80}$Zr, $^{96}$Zr, and $^{100}$Zr are similar
to what we find for $^{74}$Kr and $^{110}$Zr: changing $a_{\text{surf}}$
shifts very deformed prolate and oblate states, but does not affect much
the shape of the energy curves for $| \beta_{20} | \lesssim 0.15$, nor alter
the relative order of the coexisting minima. Both $^{80}$Zr and $^{100}$Zr
have a spherical ground state for all SLy5sX.

Altogether, this indicates that the value of the surface energy coefficient
$a_{\text{surf}}$ is an important ingredient for the correct description of
the evolution of shapes in the Zr region, with a preference for low values
like the one of SLy5s1. Other ingredients of the model such as details of
shell structure have to be precisely controlled as well, however, and
modifications of the SLy5sX in that respect are necessary.

%-----------------------------------------------------------------------

\subsection{Exotic deformation}

In the previous sections we discussed the impact of $a_{\rm surf}$ on
the ground and excited states whose shapes are dominated by
quadrupole deformations. Indeed, deformations of even
multipolarity, $\beta_{2}, \beta_{4}, \beta_{6}, \ldots$, are relevant in
virtually every region of the nuclear chart. Moments of odd multipolarity,
in particular $\beta_{30}$ and $\beta_{32}$, are in general less important
and in most cases take a zero  value for ground states calculated at
the mean-field level. Nevertheless, there are specific regions where
this is not the case. Axial octupole deformation $\beta_{30}$ plays a role for
nuclei around $^{226}$Ra, whose fission barrier has already been discussed
in Section~\ref{sec:fission}, or the region around $^{144}$Ba \cite{Bucher16}.
The possible role of dominant $\beta_{32}$ deformations in specific regions
of the chart of nuclei is also discussed in the literature~\cite{Dudek02}.

The appearance of regions of octupole deformation can be
associated with the presence of single-particle levels of opposite parity
near the Fermi energy that are mixed by the octupole deformation in such
a way that the density of levels near the Fermi energy is significantly
reduced.

Unlike quadrupole deformation of the intrinsic states, that manifests itself
through rotational bands and that can be directly measured for excited states
with angular momentum larger than $1/2$, indications for octupole
deformation are always indirect: the expectation value of a parity-odd
operator is zero in experiment. The presence of static intrinsic
$\beta_{30}$ deformation can be deduced from
rotational bands with a characteristic pattern of levels with alternating
parity that exhibit strong $E1$ and $E3$ transitions \cite{Butler96,Butler16}.

Like the calculations of fission barriers in Section~\ref{sec:fission}
the calculations discussed in this subsection have been carried out
by breaking parity, but conserving $z$~signature and $y$~time simplex of
the single-particle states, which introduces two plane reflection symmetries
of the local density $\rho(\vec{r})$ \cite{RyssensPhd}.

%- - - - - - - - - - - - - - - - - - - - - - - - - - - - - - - - - - - - -
%
\subsubsection{Axial octupole deformation}
\label{sec:octupole}

\begin{figure}[t!]
\includegraphics[width=7.3cm]{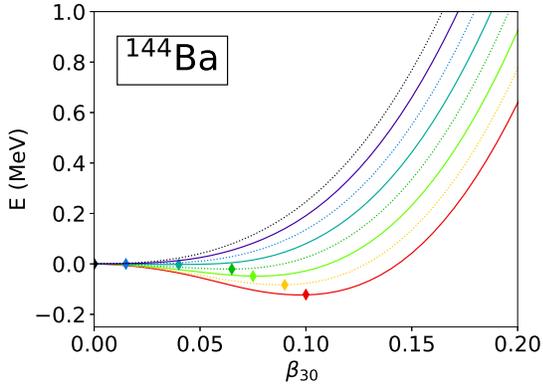}
\caption{\label{fig:B30Ba144}
Deformation energy of $^{144}$Ba relative to the respective
axial reflection-symmetric quadrupole-deformed saddle point as a function of
$\beta_{30}$ for the SLy5sX parametrizations indicated with the same
color code as in Fig.~\ref{fig:Pufission}. The minimum of each energy
curve is marked by a filled diamond.
}
\end{figure}

As a first example, we consider the medium-mass nucleus $^{144}$Ba, for which
recent experimental data from Coulomb excitation indicate static octupole
deformation~\cite{Bucher16}. Figure~\ref{fig:B30Ba144} shows the deformation
energy of this nucleus as a function of the axial octupole moment $\beta_{30}$
with diamonds indicating the position of the respective minima. As these
configurations are also quadrupole deformed, $\beta_{30} = 0$ corresponds to
an axial prolate deformed saddle point of the complete deformation energy
surface. The octupole deformation
of the minimum, which is quite appreciable for SLy5s1 with
$\beta_{30} \sim 0.15$, diminishes with increasing $a_{\text{surf}}$.
Simultaneously, the minimum becomes more shallow, until it vanishes for
SLy5s7 and SLy5s8: for these, the mean-field minimum is reflection
symmetric.

Assuming a rigid axial rotor, the reduced $E2$ matrix element
$\langle 2^+ || \mathcal{M}(E2) || 0^{+} \rangle = 1.024^{+17}_{-22} \, e \text{b}$
obtained in Ref.~\cite{Bucher16} can be identified with
$\langle Q_{20,p}\rangle$. From this, one obtains a
value of $\beta_{2,p} = 0.197^{+3}_{-4}$ for the quadrupole deformation of the
protons defined through Eq.~\eqref{equ:betalm:p}. This value compares very
well with the calculated values for $\beta_{2,p}$ that fall into the range
between 0.198 (SLy5s1) and 0.196 (SLy5s8).

Identifying the measured reduced $E3$ matrix element
$\langle 3^- || \mathcal{M}(E3) || 0^{+} \rangle
= 0.65^{+17}_{-23} \, e \text{b}^{3/2}$ along the same lines with
$\langle Q_{30,p}\rangle$, Eq.~\eqref{equ:betalm:p} leads to a value of
$\beta_{30,p} = 0.195^{+51}_{-69}$ for the proton octupole deformation. It
is appreciably larger than the largest octupole deformation of about
$\beta_{30,p} \simeq 0.1$ obtained in the calculations (for SLy5s1).
As discussed in Ref.~\cite{Bucher16}, the same underestimation of data
is also found when comparing with published $\beta_{30,p}$ values from
earlier calculations of $^{144}$Ba with a large variety of EDFs.

Extrapolating the behavior of the energy curves of Fig.~\ref{fig:B30Ba144}
beyond the range covered by the SLy5sX, it is clear that the empirical
octupole deformation cannot be reached
through just further reduction of $a_{\text{surf}}$.
For SLy5s1, the energy gain with respect to the reflection-symmetric saddle
is 200~keV, while for all other parametrizations the surface is even
softer with respect to octupole deformation. This is to be contrasted with
quadrupole deformation, which for this nucleus brings an energy gain of
several MeV. The difference in scale makes octupole deformation more
elusive and fragile than quadrupole deformation, and also more sensitive
to details of shell structure and also pairing correlations.

Octupole correlations become more pronounced, however, when projecting the
reflection-asymmetric mean-field states on parity~\cite{Robledo11}, thereby
improving the agreement with experiment~\cite{Bernard16} for this nucleus.

\begin{figure}[t!]
\includegraphics[width=9cm]{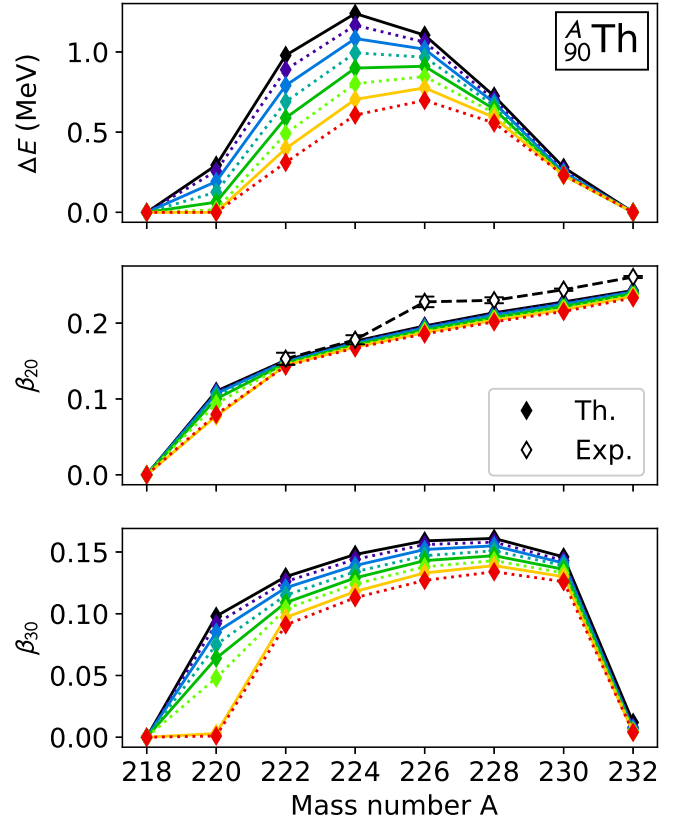}
\caption{\label{fig:ThOct}
Energy gain $\Delta E$ of the nuclear ground state from reflection-asymmetric
shape degrees of freedom, as well as quadrupole and octupole deformation
of even-even Th isotopes as obtained from calculations with the SLy5sX
parametrizations and represented with the same color code as used in
Fig.~\ref{fig:Pufission}. Experimental data for charge quadrupole deformations
$\beta_{20,p}$ taken from \cite{Raman01} and \cite{Schuler86} ($^{224}$Th)
are given for comparison.
}
\end{figure}

As an example from the $A \approx 220$ mass region, Fig.~\ref{fig:ThOct}
shows the energy gain from reflection-asymmetric shape degrees of freedom,
as well as the quadrupole and octupole deformations of the minimum, for
even-even Th isotopes between $^{214}$Th and $^{232}$Th. As in some cases
non-constrained HFB+LN calculations did not converge to the minimum
of the energy surfaces because of the non-variational character of the
LN procedure, the results were obtained from interpolation of energy
surfaces constructed around the minimum.
For $^{214}$Th, $^{216}$Th (both not shown), and $^{218}$Th, the
lowest mean-field configuration is in fact spherical because of the
proximity to the $N=126$ shell closure. From $^{220}$Th onwards, octupole
and quadrupole deformation set in simultaneously. For the transitional
$^{220}$Th, a very shallow octupole-deformed minimum is found for all
parametrizations but SLy5s7 and SLy5s8, and that in the most favorable
case of SLy5s1 is 294~keV below the reflection-symmetric saddle.

The heavier isotopes up to $^{228}$Th exhibit a much more pronounced
asymmetric minimum, with an energy gain of the order of 1~MeV
for SLy5s1, which is significantly larger than what we found for $^{226}$Ra
and $^{144}$Ba. Beyond $^{228}$Th, the octupole
deformation is quickly decreasing again, while the quadrupole deformation
continues to grow. The size of octupole deformation is
correlated to $a_{\text{surf}}$: for all strongly octupole-deformed
Th isotopes shown in Fig.~\ref{fig:ThOct} the value of $\beta_{30}$ for
the calculated minimum significantly decreases with increasing
$a_{\text{surf}}$. The values obtained with SLy5s8 are typically $30\%$
smaller than those from SLy5s1. The same correlation with $a_{\text{surf}}$
can also be found for the quadrupole deformation $\beta_{20}$, but on a
much smaller scale that is not significant. With the exception of an
anomaly for $^{226}$Th, all SLy5sX describe very well the size and global
trend of the experimental $\beta_{20,p}$.

From the available experimental information for rotational bands in these
nuclei it has been concluded that the spectrum of $^{220}$Th cannot be
interpreted in terms of an octupole-deformed rigid rotor~\cite{Reviol14}.
Alternating parity bands are observed starting from $^{222}$Th
onwards \cite{Cocks99}. For $^{222}$Th, $^{224}$Th, and $^{226}$Th, the
pattern can be interpreted in terms of a rigid octupole rotor, whereas the
data for $^{230}$Th and $^{232}$Th suggests that at low spin the octupole
deformation is vibrational. The bands of the latter resemble the one of
an octupole-deformed rotor only at higher spin, while $^{228}$Th is
transitional in between the two regimes \cite{Cocks99}. Up to $^{232}$Th,
our findings are compatible with the data, keeping in
mind that, like in the case of transitional quadrupole-deformed nuclei,
it is not obvious to deduce how the particular spectroscopic features
of $^{220}$Th on the one hand and the vibrational character of $^{230}$Th
and $^{232}$Th on the other hand should be reflected by their mean-field
deformation energy surface. In any event, the very shallow minima found
for these three
nuclei (as indicated by their small energy gain) suggest that a mere
mean-field calculation might not be sufficient to describe states in this
nucleus. In particular the energy surfaces of $^{230}$Th and $^{232}$Th
have a long leveled valley in $\beta_{30}$ direction at almost constant
$\beta_{20}$ \cite{Ryssens18x} that could indeed lead to large-amplitude
octupole vibrations.

For the isotopes that can be interpreted in terms of a static octupole rotor,
there are no available data for $B(E3)$ transitions to the ground
state.\footnote{Putting the $B(E3,0^+ \to 3^-)$ values of the
\textit{vibrational} $^{230}$Th and $^{232}$Th reported in
Ref.~\cite{McGowan74} into the expression for $\beta_{30,p}$ of an
octupole-deformed \textit{rotor}, one obtains $\beta_{30,p} = 0.094(29)$
for $^{230}$Th and $\beta_{30,p} = 0.085(28)$ for $^{232}$Th, respectively.}

Other forms and parametrizations of the nuclear EDF overall agree on the
octupole deformation of Th isotopes in this mass region, but might differ
in details \cite{Butler16}.
In calculations using the D1S and D1N parametrizations of the Gogny force,
the onset of octupole deformation in the mean-field ground states is also
found for $^{220}$Th, whereas for D1M the lightest octupole-deformed
isotope is $^{222}$Th~\cite{Robledo11}. The DD-PC1 and NL3
parametrizations of relativistic EDFs predict octupole deformation only
beginning with $^{224}$Th~\cite{Agbemava16} or even
$^{226}$Th~\cite{Nomura13}, respectively.

At the mean-field level, SLy5s1 with its low $a_{\text{surf}}$ gives the
most pronounced octupole deformation, which is also the parametrization
that tends to agree best with the data discussed up to now.
Further discussion of the
structure of $^{222}$Th, including its rotational band, as obtained from
calculations with SLy5s1 can be found in Ref.~\cite{Ryssens18a}. A very
detailed study of the structure of even and odd Th isotopes also using
SLy5s1 will be presented elsewhere~\cite{Ryssens18x}.

%- - - - - - - - - - - - - - - - - - - - - - - - - - - - - - - - - - - - -
%
\subsubsection{Non-axial octupole deformations}

In Section~\ref{sec:fission}, we saw that non-axial octupole deformations
$\beta_{32}$ can lower the static fission path around the saddle point,
as has been reported earlier, for example in Refs.~\cite{Schunck14,Lu14a}.

\begin{figure}[t!]
\includegraphics[width=8cm]{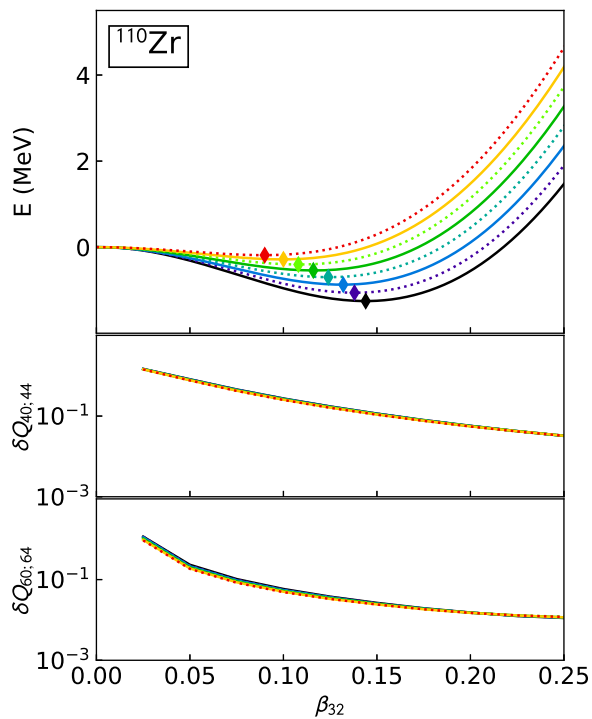}
\caption{\label{fig:Zr110tetra}
Deformation energy of $^{110}$Zr as a function of the non-axial octupole
deformation $\beta_{32}$ from HF calculations without pairing correlations
using the SLy5sX parametrizations as indicated.
The lower panels show the absolute value of the relative deviation between
higher-order multipole deformations defined as
$\delta Q_{40:44} \equiv \big|  (Q_{40} + \sqrt{14/5} \; Q_{44}) /
Q_{40} \big|$
and
$\delta Q_{60:64} \equiv \big| ( Q_{60} - \sqrt{2/7} \; Q_{64}) /
Q_{60} \big|$ that characterize the closeness to an octahedral solution
(see text). All configurations along the curve have been constrained to have
$\beta_{20} = \beta_{22} = 0$ in order to prevent the code from converging to
the lower-lying quadrupole-deformed minimum of the energy surfaces displayed
in Fig.~\ref{fig:Zr110:beta:gamma}.
}
\end{figure}

The relevance of $\beta_{32}$ for low-lying states at normal deformation
was originally discussed to characterize cluster structures in light
nuclei~\cite{Onishi71}. It is the leading deformation degree of freedom
characterizing tetrahedral and octahedral shapes~\cite{Dudek02,Dudek07}.
Similar to quadrupole deformation, such symmetries are predicted to give
rise to substantial deformed shell gaps at specific particle numbers and,
hence, might be present in nuclear ground states or excited states at
low excitation energy.

In the expansion of the surface of a liquid drop along the lines of
Eq.~\eqref{eq:dropexpansion} that includes all spherical harmonics,
octahedral shapes are characterized by finite $\alpha_{32}$,
$\alpha_{40} = - \sqrt{14/5} \; \alpha_{44}$, and
$\alpha_{60} =   \sqrt{2/7} \; \alpha_{64}$, with all other $\alpha_{\ell m}$
equal to zero up to $\ell = 6$ \cite{Dudek07}. Note that such shape
is also invariant under the tretrahedral point group \cite{Dudek07}.
For the same reasons that lead to a
difference between $\alpha_{2 0}$ and $\beta_{2 0}$ discussed above, however,
the deformations $\beta_{\ell m}$ of Eq.~\eqref{equ:betalm} calculated from
multipole moments are not equal to the $\alpha_{\ell m}$ such that the
relations between the $\alpha_{\ell m}$ of same $\ell$ cannot be expected
a priori to also hold exactly for the $\beta_{\ell m}$.

The neutron-rich $^{110}$Zr is one of the the most detailed studied
candidates for tetrahedral
deformation~\cite{Dudek02,Schunck04,Zberecki09a,Tagami15,Zhao17}. Recent
experimental evidence~\cite{Paul17} has invalidated
the prediction for the ground state band, although excited bands could
still exhibit tetrahedral or octahedral character.

Figure~\ref{fig:Zr110tetra} shows the energy curve of $^{110}$Zr as a
function of $\beta_{32}$. The quadrupole deformations $\beta_{20}$
and $\beta_{22}$ are constrained to be zero, such that $\beta_{32} = 0$
corresponds to a spherical shape. Unlike all other calculations
presented in this study, these were performed at the HF level
neglecting pairing correlations. In HFB+LN calculations using the same
pairing interaction as above, there is no deformed minimum for any of the
SLy5sX; instead, the curves are slowly rising with $\beta_{32}$. By
contrast, HF calculations yield a very shallow minimum for all SLy5sX.
Like in the case of the axial octupole deformation of $^{144}$Ba and
the Th isotopes, with decreasing $a_{\text{surf}}$ the minima become
more pronounced in both width and depth.

The lower two panels show the relative deviation of
$-\sqrt{14/5} \; \beta_{4 4}$ from $\beta_{4 0}$ and of
$\sqrt{2/7} \; \beta_{6 4}$ from $\beta_{6 0}$. The relations between the
surface moments $\alpha_{\ell m}$ mentioned above are reasonably well
respected.
Up to numerical noise, the other non-constrained low-order deformations take
a value of zero,
$\beta_{30} = \beta_{42} = \beta_{5 \ell} = \beta_{62} = \beta_{66}$,
indicating that the shapes along the energy curve exhibit indeed
octahedral symmetry.
For SLy5s1, the actual multipole deformations at the minimum are
$\beta_{32} = 0.15$, $\beta_{40} = -0.027$, $\beta_{44} = 0.014$,
$\beta_{60} = 0.019$ and $\beta_{64} = 0.037$.

%
%=========================================================================
%
\section{Conclusions}
\label{sec:conclusion}

We have studied the correlation of the value of the surface energy coefficient
$a_{\text{surf}}$ with observables characterizing deformation phenomena in
atomic nuclei. To that end, we performed calculations with the recent
SLy5sX parametrizations of the standard Skyrme EDF that were each adjusted
within the same protocol with a constraint on a systematically varied
value of $a_{\text{surf}}$. Going from SLy5s1 with the lowest
$a_{\text{surf}}$ to SLy5s8 with the highest value covers the range of
$a_{\text{surf}}$ typically found for widely-used Skyrme parametrizations.

Using a family of fits for which all other properties are as similar
as possible is crucial for such study. This is particularly pertinent
with regard to shell effects. Indeed, the complicated topography of
deformation energy surfaces with multiple deformed minima and saddles
in between is generated by shell effects, with the surface energy only
providing a smooth background.

The main conclusions concerning the description of properties of finite
nuclei that can be drawn are as follows:
\begin{enumerate}[(i)]
\item
As expected, the deformation energy of highly-deformed configurations is
clearly correlated to $a_{\text{surf}}$. For the saddle points
of very wide fission barriers of heavy nuclei, the difference between
what is obtained with SLy5s1 and SLy5s8 can amount to as much as 10~MeV.

\item
The description of fission barrier heights of nuclei in the $A \approx 240$
actinide and neutron-deficient $A \approx 180$ Hg region improves dramatically
when reducing $a_{\text{surf}}$, with a clear preference for the SLy5s1
parametrization.

\item
The performances of the SLy5sX for the barrier height of $^{240}$Pu and
$^{180}$Hg, two nuclei with very different asymmetry $I$, are not the same.
When the barrier height of $^{240}$Pu is correctly described, then the one
of $^{180}$Hg is largely overestimated. One possible explanation is
that the surface symmetry energy coefficient $a_{\text{ssym}}$, which takes
very similar values for all SLy5sX, needs fine-tuning too, such that the
effective surface energy coefficient $a_{\text{surf,eff}}(I)$ decreases
less quickly with asymmetry $I$. This, however, would require a further
reduction of $a_{\text{surf}}$ below the SLy5s1 value.

\item
The clear correlation between barrier height and  $a_{\text{surf}}$ fades
away when going to superheavy nuclei in the $Z \simeq 110$ region. These
systems are characterized by a vanishing liquid-drop fission barrier, such
that solely the details of shell structure determine the fission barrier.
For these systems, all SLy5sX give very similar results.

\item
Results for the excitation energy of the superdeformed minimum of nuclei in
the $A \approx 190$ region do not follow the same trends as the fission barrier
heights: for some nuclei such as $^{190}$Hg it is even underestimated
by all SLy5sX parametrizations. This has to be contrasted with the fission
barrier of $^{180}$Hg, that is overestimated by all SLy5sX
parametrizations, in particular those with high $a_{\text{surf}}$.
In general, the excitation energy of the superdeformed states increases
too rapidly with asymmetry~$I$. As the fission isomer of the much more
asymmetric $^{240}$Pu is reasonably described by the SLy5s1
parametrization that also fairly describes the barrier height, the
difficulties to describe the known superdeformed states of Hg and Pb nuclei
have to be a local particularity of the $A \approx 190$ region.
The most likely explanation is that the modeling
of these states is compromised by deficiencies in the description of shell
effects and their dependence on $N$, $Z$, and deformation. Improvements of
this aspect of nuclear EDFs could also slightly alter the conclusions about
fission barrier heights drawn above, but are unlikely to be achievable within
the present standard form of the Skyrme EDFs.

\item
At normal deformation, the changes in surface energy when going from one
SLy5sX parametrization to another are naturally smaller and also often
masked by simultaneous small changes in shell effects, in particular
for oblate states. The relative energy between coexisting minima
is less strictly correlated to $a_{\text{surf}}$ as is the case at larger
deformation. None of the SLy5sX parametrizations provides a correct
description of shape coexistence and shape evolution in the region of
Kr and Zr isotopes at the mean-field level, a problem they share with
many other nuclear EDFs. Still, in most cases
the parametrizations with low $a_{\text{surf}}$ value are much closer
to experiment than those with large $a_{\text{surf}}$.

\item
For the majority of cases that we have studied, the SLy5sX parametrizations give
very similar values for the quadrupole deformations of a given normal-deformed
or superdeformed minimum in the energy surface. Differences are on the few
percent level. In many cases, a SLy5sX parametrization
with smaller $a_{\text{surf}}$ gives slightly higher $\beta_{20}$ than
a parametrization with higher $a_{\text{surf}}$, but that is clearly
not a general rule. In any event, the differences between
the parametrizations for quadrupole deformations
are rarely significant. In most cases these values also agree very well
with data from electromagnetic transition matrix elements in the yrast
bands built on the deformed configuration in question. Similarly,
predictions for higher multipole deformations with even $\ell$ are
very similar for the majority of cases.

\item
The situation is quite different for nuclei in regions where octupole
deformation, either axial or non-axial, plays a role for the ground state.
The octupole deformation of the minimum becomes significantly more
pronounced when reducing $a_{\text{surf}}$, which also tends to improve
agreement with (indirect) experimental data. In turn, in some cases such
minima can disappear completely when using a parametrization with large
$a_{\text{surf}}$. The available data for octupole deformed nuclei also
show a clear preference for parametrizations with low  $a_{\text{surf}}$.
Remaining deviations, however, indicate that there is also need for
improvement of other properties of the nuclear EDF, most importantly
details of shell structure that is at the origin of these minima.

\end{enumerate}
It is important to note that all time-even terms in the EDF contribute
to $a_{\text{surf}}$, not just the gradient terms. As a consequence,
in EDF-based methods $a_{\text{surf}}$ is intertwined with the properties
of infinite nuclear matter and shell structure. This has to be contrasted
with macroscopic-microscopic approaches, where these three aspects of the
model can be independently adjusted. For the SLy5sX
parametrizations, the total size of all contributions to $a_{\text{surf}}$
change when going from one parametrization to another:
\begin{enumerate}[(i)]
\item
The terms in the EDF that contribute to the energy per particle of infinite
nuclear matter contribute to about half the total value of $a_{\text{surf}}$.
As a consequence, the self-consistency of the protocol for parameter
adjustment strongly correlates $a_{\text{surf}}$ with $a_{\text{vol}}$ and
$a_{\text{sym}}$.
This correlation is such that the
gross properties of two-neutron and two-proton separation energies
are the same for all the SLy5sX parametrizations, in spite of their quite different symmetry
energy coefficients.

\item
Because of a sizable contribution from the spin-orbit interaction to
$a_{\text{surf}}$, which also has the opposite sign of the other large
contributions, spin-orbit splittings of the SLy5sX parametrizations are
correlated to $a_{\text{surf}}$. This has some visible impact on
shell structure, in particular the position of high-$j$ intruder levels
and the amplitude of the variation of shell effects in the deformation energy
surfaces.
\end{enumerate}
We note in passing that we did not find any significant correlation
between the density profile of spherical nuclei obtained with the SLy5sX
parametrizations and the value of their $a_{\text{surf}}$.

The construction of the SLy5sX parametrizations~\cite{Jodon16} is
part of ongoing efforts to improve the fit protocol of nuclear EDFs.
The new element that we have thoroughly studied concerns the control of the
deformation properties of EDFs. It confirms that superdeformed
states and fission
barriers are, as expected, sensitive to the fine-tuning of the
surface energy coefficient. At the mean-field
level of modeling, results for highly-deformed states obtained with SLy5s1
are clearly superior to those obtained with the majority of other Skyrme
parametrizations~\cite{Jodon16,Nikolov11a}, which have usually been adjusted
without any regard to the surface energy.
Compared to earlier widely-used parametrizations of the
Skyrme EDF that reach similar quality for highly-deformed states, as
SkM* \cite{Bartel82}, UNEDF1 \cite{unedf1}, and UNEDF2 \cite{unedf2},
SLy5s1 has several advantages. It performs much better for isotopic
and isotonic trends of binding energies than SkM* (compare, for example,
Fig.~\ref{fig:s2n:hg} with results presented in Ref.~\cite{Heenen98}),
whose deficiencies in that respect were already pointed out in the 
original paper \cite{Bartel82}. Also,
unlike the UNEDF1 and UNEDF2 parametrizations that only define the
time-even terms in the functional, SLy5s1 can be used without ambiguities
to calculate time-odd terms in situations where time-reversal symmetry is
broken, such as the calculation of odd- and odd-odd nuclei or the calculation
of rotational bands. Therefore, SLy5s1 will be our parametrization of choice
for future studies of the properties of heavy nuclei.

The full set of SLy5sX parametrizations can also be used for further
studies of correlations between the surface energy coefficient and other 
properties of nuclei not addressed here. It presents the opportunity to
complement studies of correlations between observables in finite nuclei
and properties of infinite nuclear matter that can be carried out with
the SV-based parametrizations of Ref.~\cite{Klupfel09} or the 
SAMi-based parametrizations of Refs.~\cite{RocaMaza13,Cao15}.
Indeed, only families of fits that systematically vary a single property
of the EDF allow for controllable correlation analyses.

The results presented here suggest that a simultaneous adjustment
of the surface and surface symmetry energy coefficients will be needed
for further improvement. They also show, however, that it is 
not sufficient to fine tune only the surface energy. A better control of
shell effects, which are at the origin of deformed minima, is equally
important, in particular for the description of states with exotic shapes.
The deficiencies of nuclear EDFs for single-particle spectra also
concern many other observables. Earlier studies, however, indicate that
it is unlikely that they can be systematically resolved within the current
form of nuclear EDFs \cite{unedf2,Lesinski07a}.

One has to note that the value of $a_{\text{surf}}$ is not model
independent. Its extraction from an EDF depends on the model that will be used
to calculate nuclei \cite{Jodon16}. Therefore,
its ``best value'' as defined here is valid for parametrizations
designed for mean-field calculations. It has to be redefined if correlations beyond 
mean-field are introduced, as corrections for spurious rotational motion or mixing of
mean-field configurations.
The same dependence on the model can be expected for the surface symmetry energy coefficient
$a_{\text{ssym}}$. A comparison of several frequently-used procedures for
its calculation is presently underway \cite{Meyer18x}, with the goal of
finding a computationally-friendly way to constrain it during
parameter fits.

%==========================================================================
%==========================================================================
%==========================================================================

\section*{Acknowledgments}

We thank J.~Bartel for an illuminating discussion on the construction
of the SkM* parametrization. This work was supported by the
French Centre national de la recherche scientifique (CNRS) through PICS Grant 
No. 6949 through PICS Grant No.~6949, the University of Jyv{\"a}skyl{\"a} within
the FIDIPRO program, and the IAP Belgium Science Policy (Brix network P7/12).
The computations were performed using HPC resources from the
Consortium des {\'E}quipements de Calcul Intensif (C{\'E}CI),
funded by the Fonds de la Recherche Scientifique de Belgique
(F.R.S.-FNRS) under Grant No.~2.5020.11, and the CCIN2P3 of the CNRS.

%==========================================================================
%==========================================================================
%==========================================================================

\end{document}